%% file: Bohnet_pennSqueeze_15.tex
\newcommand{\ket}[1]{\ensuremath{\left|  #1 \right\rangle}}

\newcommand{\Be}{\ensuremath{^9$Be$^+}}
\newcommand{\Jbar}{\ensuremath{\bar{J}}}
\newcommand{\vecS}{\ensuremath{ |\langle\vec{S}\rangle|}}

\documentclass[12pt]{article}
\usepackage{graphicx}
\usepackage{fixltx2e}
\usepackage{amsmath}
\usepackage{amssymb}
\usepackage{bm}
\usepackage{moreverb}
\usepackage{setspace}

\newcommand{\be}{\begin{equation}}
\newcommand{\ee}{\end{equation}}
\newcommand{\bwt}{\begin{widetext}}
\newcommand{\ewt}{\end{widetext}}
\newcommand{\bea}{\begin{eqnarray}}
\newcommand{\eea}{\end{eqnarray}}

\newcommand{\sinc}{\mathrm{sinc}}

\usepackage{scicite}

\usepackage{times}

\newenvironment{sciabstract}{%
\begin{quote} \bf}
{\end{quote}}

\newcounter{lastnote}

\title{Title: Quantum spin dynamics and entanglement generation with hundreds of trapped ions}

\author
{Authors: Justin G. Bohnet$^{1\ast}$,  Brian C. Sawyer$^{2}$, Joseph W. Britton$^{1}$,\\  Michael L. Wall$^{3}$, Ana Maria Rey$^{4}$, Michael Foss-Feig$^{5}$, John J. Bollinger$^{1}$\\
\\
\normalsize{Affiliations: $^{1}$NIST, Boulder, Colorado 80305, USA},
\normalsize{$^{2}$GTRI, Atlanta, Georgia 30332, USA}\\
\normalsize{$^{3}$JILA, NIST and University of Colorado, Boulder, Colorado, 80309, USA}\\
\normalsize{$^{4}$JILA, NIST and Department of Physics, University of Colorado, Boulder, Colorado, 80309, USA}\\
\normalsize{$^{5}$Joint Quantum Institute and NIST, Gaithersburg, Maryland, 20899, USA}
\\
\normalsize{$^\ast$To whom correspondence should be addressed; E-mail:  justin.bohnet@nist.gov}
}

\date{}

\begin{document} 


\baselineskip24pt


\maketitle 


\begin{sciabstract}
Abstract: Quantum simulation of spin models can provide insight into complex problems that are difficult or impossible to study with classical computers. Trapped ions are an established platform for quantum simulation, but only systems with fewer than 20 ions have demonstrated quantum correlations. Here we study non-equilibrium, quantum spin dynamics arising from an engineered, homogeneous Ising interaction in a two-dimensional array of \Be{} ions in a Penning trap. We verify entanglement in the form of spin-squeezed states for up to 219 ions, directly observing 4.0$\pm$0.9 dB of spectroscopic enhancement. We also observe evidence of non-Gaussian, over-squeezed states in the full counting statistics. We find good  agreement with ab-initio theory that includes competition between entanglement and decoherence, laying the groundwork for simulations of the transverse-field Ising model with variable-range interactions, for which numerical solutions are, in general, classically intractable.


\end{sciabstract}


\paragraph*{One Sentence Summary:} We report experimental measurements of open-system Ising spin dynamics in two-dimensional arrays of more than 200 trapped ions, demonstrating entanglement by observing spin-squeezing and acquiring full counting statistics, both of which show quantitative agreement with theory that accounts for decoherence.

\paragraph*{Main Text:} Quantum simulation, where one well-controlled quantum system emulates another system to be studied, anticipates solutions to intractable problems in fields including condensed-matter and high-energy physics, cosmology, and chemistry, before the development of a general purpose quantum computer\cite{Feynman1982,Hauke2012,Georgescu2014}.
Of particular interest are simulations of the transverse-field Ising spin model\cite{Elliott1970}, described by the Hamiltonian 
\begin{equation}
	\hat{H}_T = \hat{H}_I + \hat{H}_B\,
	\label{eqn:TransverseIsing}
\end{equation}
\begin{equation}
	\hat{H}_I = \frac{1}{N}  \sum_{i<j}^N J_{i,j} \hat{\sigma}^z_i \hat{\sigma}^z_j \,, \,\, \hat{H}_B = \sum_i^N B_x \hat{\sigma}^x_i \,,
	\label{eqn:Ising}
\end{equation}
\noindent where $N$ is the number of spins, $J_{i,j}$ parameterizes the spin-spin interaction, $B_x$ parameterizes a transverse magnetic field, and $\hat{\sigma}^z$, $\hat{\sigma}^x$ are Pauli spin matrices. 
A quantum simulation of $\hat{H}_T$  could illuminate complex phenomena in quantum magnetism, including quantum phase transitions, many-body localization, and glassiness in spin systems \cite{Binder1986,Nandkishore2015,Belitz2005,Sachdev2007}, and clarify whether quantum annealing can provide a speed-up for solving hard optimization problems\cite{Ronnow2014,Lucas2014}.

Ensembles of photons, ions, neutral atoms, molecules, and superconducting circuits are all developing as quantum simulation platforms\cite{Georgescu2014}.
For example, a variety of quantum spin models have been realized with large ensembles of neutral atoms\cite{Simon2011,DePaz2013,Fukuhara2015,Brown2015} and molecules\cite{Yan2013}, using contact or dipolar interactions in optical lattices and using infinite-range interactions mediated by photons in optical cavities\cite{Leroux2010c}.
Trapped-ion quantum simulators can implement $\hat{H}_T$\cite{Porras2004,Jurcevic2014,Richerme2014} and have a number of advantages over other implementations, such as high-fidelity state preparation and readout, long trapping and coherence times, and strong, variable-range spin-spin couplings.
To date, trapped-ion simulators have been constrained to of order 20 spins\cite{Islam2013,Jurcevic2014}, where classical numerical simulation remains tractable, but substantial engineering efforts are underway to increase the number of ions by cryogenically cooling linear traps and 2D surface-electrode traps\cite{Schmied2009,Bruzewicz2015}.


Penning traps have emerged as an alternative approach to performing quantum simulations with hundreds of ions\cite{Britton2012,Sawyer2012,Wang2013}.
Laser-cooled ions in a Penning trap self assemble into two-dimensional triangular lattices and are amenable to the same high-fidelity spin-state control, long trapping times, and generation of transverse-field Ising interactions as ions in linear Paul traps. 
Previous work in Penning traps demonstrated control of the collective spin\cite{Biercuk2009} and benchmarked the engineered, variable-range Ising interaction in the mean-field, semi-classical limit\cite{Britton2012,Sawyer2012,Wang2013}.
However, for a simulator of quantum magnetism to be trusted, quantum correlations generated by the Ising interaction must be observed and understood.
For large, trapped-ion simulators, this benchmarking requires a detailed accounting of many-body physics in an open quantum system.
This is both a challenge and an opportunity, as existing numerical methods for computing strongly correlated dynamics of open quantum systems are, in general, inadequate to model large ensembles.
 
Here, we observe and benchmark entanglement in hundreds of trapped ions generated with engineered Ising interactions in a 2D array of \Be{} ions in a Penning trap.
To enable efficient theoretical computation of the spin dynamics\cite{SM_sci}, we perform experiments with a homogenous Ising interaction and without simultaneous application of the transverse field $B_x$.
We use global spin observables, such as collective magnetization, to study quantum correlations without needing to perform full state tomography on the ensemble\cite{Kastner2011,Worm2013,Foss-Feig2013,Hazzard2014}.
For each experimental result, we verify that a solution of the full quantum master equation gives good agreement with the data.
Specifically, we perform three measurements that characterize the quantum dynamics.
We observe an $N$-dependent depolarization of the collective spin, distinguishing the destruction of correlations caused by decoherence from the coherent depolarization caused by the spin-spin interaction. 
We measure the modification of the collective spin variance arising from correlations between the spins, and verify that entanglement survives the decoherence using spin-squeezing as an entanglement witness\cite{Wineland1992,Kitagawa1993,Sørensen2001c}.
For longer interaction times, when spin-squeezing is absent, we observe non-Gaussian counting statistics in the collective spin state.
Comparison of the full counting statistics with numerical calculations shows that the disappearance of spin-squeezing is due to the formation of an over-squeezed state\cite{Shalm2009,Pezze2009} and not just degradation due to decoherence.

\begin{figure*}
\includegraphics[width=6.4in]{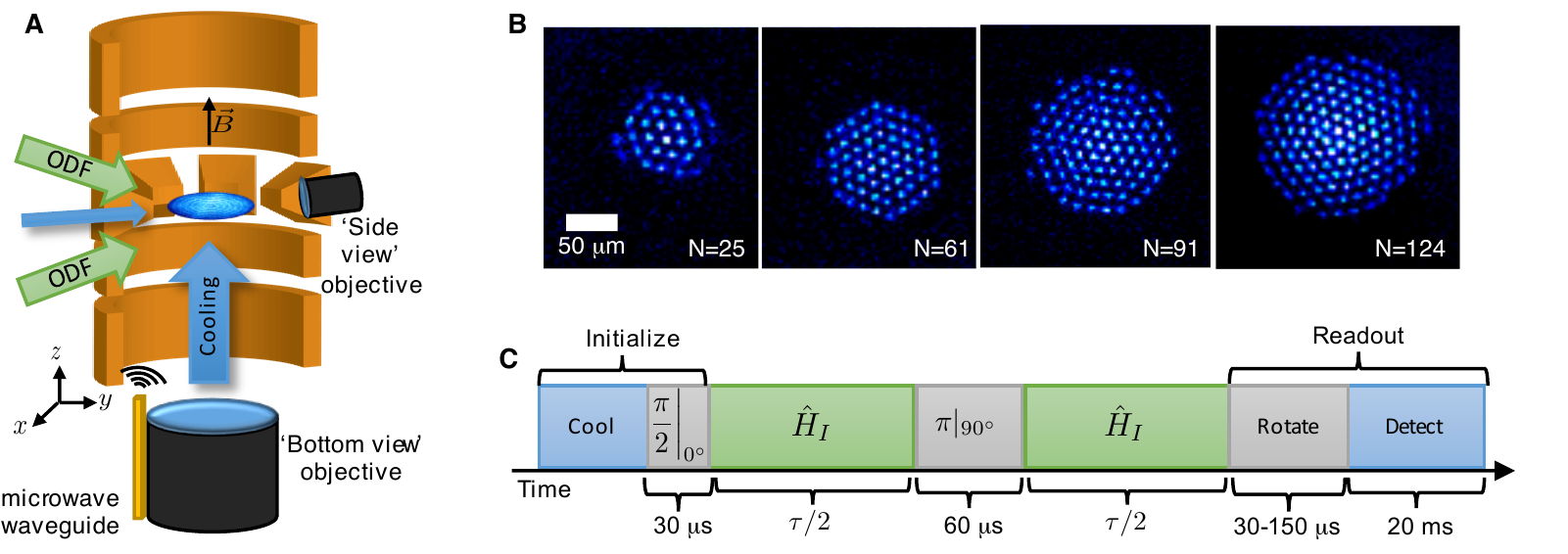}
\caption{{\bf Penning trap quantum simulator.} {\bf (A)} A cross-section illustration of the Penning trap (not to scale). The orange electrodes provide axial confinement and the rotating wall potential. The $4.5$ T magnetic field is directed along the $z$-axis. The blue disk indicates the 2D ion crystal. Resonant Doppler cooling is performed with the beams along $z$ and $y$. The spin-state dependent optical dipole force (ODF) beams enter $\pm10$ degrees from the 2D ion plane. Resonant microwave radiation for coupling ground states $\ket{\uparrow}$ and $\ket{\downarrow}$ is delivered through a waveguide. State-dependent fluorescence is collected through the pair of imaging objectives, where the bright state corresponds to $\ket{\uparrow}$. {\bf(B)} Coulomb crystal images in a frame rotating at $\omega_r$ with \Be{} ions in \ket{\uparrow}. {\bf(C)} The typical experiment pulse sequence, composed of cooling laser pulses (blue), microwave pulses (grey), and ODF laser pulses (green). Cooling and repumping initialize each ion in $\ket{\uparrow}$, then a microwave $\pi/2$ pulse prepares the spins along the $x$-axis. Suddenly switching on $\hat{H}_I$ initiates the non-equilibrium spin dynamics. The microwave $\pi$ pulse implements a spin-echo, reducing dephasing from magnetic field fluctuations and ODF laser light shifts. State readout consists of a final global rotation and fluorescence detection. The final microwave pulse area and phase are chosen to measure the desired spin projection.}
\label{fig:expt}
\end{figure*} 

\label{sec: apparatus}
Our experimental system consists of between 20 and 300 \Be{} ions confined to a single-plane Coulomb crystal in a Penning trap, described in Fig. \ref{fig:expt} and \cite{SM_sci}.
The trap is characterized by an axial magnetic field $|\vec{B}| = 4.45$ T and an axial trap frequency $\omega_z = 2 \pi \times 1.57$ MHz. 
A stack of cylindrical electrodes generates a harmonic confining potential along their axis.  
Radial confinement is provided by the Lorentz force from $\vec{E}\times\vec{B}$-induced rotation in the axial magnetic field.
Time varying potentials applied to eight azimuthally segmented electrodes generate a rotating wall potential that controls the crystal rotation frequency $\omega_r$, typically between $2\pi\times$ 172 kHz and $2\pi\times$ 190 kHz.

The spin-1/2 system is the $^2S_{1/2}$ ground state of the valence electron spin $\ket{\uparrow}(\ket{\downarrow}) \equiv \ket{m_s = +1/2}(\ket{m_s = -1/2})$.
In the magnetic field of the Penning trap, the ground state is split by 124 GHz. 
A resonant microwave source provides an effective transverse field, which we use to perform global rotations of the spin ensemble with a Rabi frequency of 8.3 kHz. 
The $T_2$ spin echo coherence time is 15 ms.  
Optical transitions to the  $^2P_{3/2}$ states are used for state preparation, Doppler cooling, and projective measurement\cite{SM_sci}.

\label{sec:interaction}
The Ising interaction is implemented by a spin-dependent optical dipole force (ODF) generated from the interference of a pair of detuned lasers, shown in Fig. \ref{fig:expt}a.
The ODF couples the spin and motional degrees of freedom through the interaction $\hat{H}_{ODF} = \sum_{i=1}^N F_0 \cos(\mu t)\hat{z}_i\hat{\sigma}^z_i$, where $\hat{z}_i$ is the position operator for ion $i$, $\frac{\mu}{2\pi}$ is the ODF laser beatnote frequency, and $F_0$ is the force amplitude, typically $30$ yN.
The ODF drives the axial drumhead modes of the planar ion crystal\cite{Sawyer2012,Wang2013}, generating an effective spin-spin interaction by modifying the ions' Coulomb potential energy\cite{Leibfried2003}.
Detuning $\mu$ from $\omega_z$ changes the effective range of the spin-spin interaction $J_{i,j}\propto d_{i,j}^{-a}$, where $d_{i,j}$ is the ion separation.
Although $a$ can range from 0 to 3\cite{Britton2012}, in this work we primarily drive the highest frequency, center-of-mass (COM) mode at $\omega_z$ with ODF detunings $\delta=\mu-\omega_z$ ranging from about $2\pi\times 0.5$ kHz to $2\pi\times 3$ kHz, such that $a$ varies from $0.02$ to $0.18$, respectively. 
The next closest axial motional mode frequency is more than $2\pi\times 20$ kHz lower than $\omega_z$.
Since $a \ll 1$, the Ising interaction is approximately independent of distance, resulting in a homogeneous pairwise coupling $J_{i,j} \approx \bar{J} = \frac{F_0^2}{4 M \omega_z \delta}$, where $M$ is the ion mass.

At the mean-field level, each spin precesses in an effective magnetic field determined by the couplings to other spins, described by the Hamiltonian $\hat{H}_{MF} = \sum_{j=1}^N\bar{B}_j\hat{\sigma}_j^z/2$, where $\bar{B}_j = \frac{2}{N}\sum_{i\neq j} J_{i,j}\langle \hat{\sigma}_i^z\rangle$.
We calibrate \Jbar{} through measurements of mean-field spin precession\cite{Britton2012,SM_sci}, typically finding $\Jbar{}/h$ $\leq$ 3300 Hz.
For the experiments described below, we start with all the spins initialized in an eigenstate of $\hat{\sigma}^x$ so that $\bar{B}_j = 0$.
This choice of initial condition ensures the observed physics are dominated by quantum correlations and decoherence alone.

\label{sec:state_readout}
State readout is performed using fluorescence from the Doppler cooling laser on the cycling transition\cite{SM_sci}. Ions in $\ket{\uparrow}$ fluoresce and ions in $\ket{\downarrow}$ are dark.
Global fluorescence is collected with the side view objective (Fig. \ref{fig:expt}a) and counted with a photomultiplier tube.
We calibrate the photon counts per ion using the bottom view image to count the number of ions (Fig. \ref{fig:expt}b).
From the detected photon number, we infer the bright state population $N_\uparrow$, which is equivalent to a projective measurement of $\hat{S}_z = \hat{N}_\uparrow - N/2$, where $\hat{S}_z$ is the $z$ component of the collective spin vector $\vec{S} = \frac{1}{2} \sum_i^N \left( \hat{\sigma}^x_i, \hat{\sigma}^y_i, \hat{\sigma}^z_i \right)$.  
By performing a final global rotation before measuring, we can measure the moments of any component of $\vec{S}$.  
The directly observed variance of the measurement $(\Delta S_z)^2$ is well described by the sum of two noise terms: spin noise $(\Delta S_z')^2$ and photon shot noise $(\Delta S_{psn})^2$.
Here $\Delta X$ indicates the standard deviation of repeated measurements of $\hat{X}$.
In this paper, we use the underlying spin noise $(\Delta S_z')^2 = (\Delta S_z)^2 - (\Delta S_{psn})^2$ for comparison with theory predictions, but use the directly observed variance in the measurement $(\Delta S_z)^2$ for evaluating the spin-squeezing entanglement witness.
The ratio $(\Delta S_{psn})^2/(\Delta S_z')^2$ is typically 0.13 (-8.8 dB), so the noise subtraction is small for all but the most squeezed states observed here.
Other sources of technical noise in the state readout are not significant\cite{SM_sci}.

\begin{figure}
\includegraphics[width=4.8in]{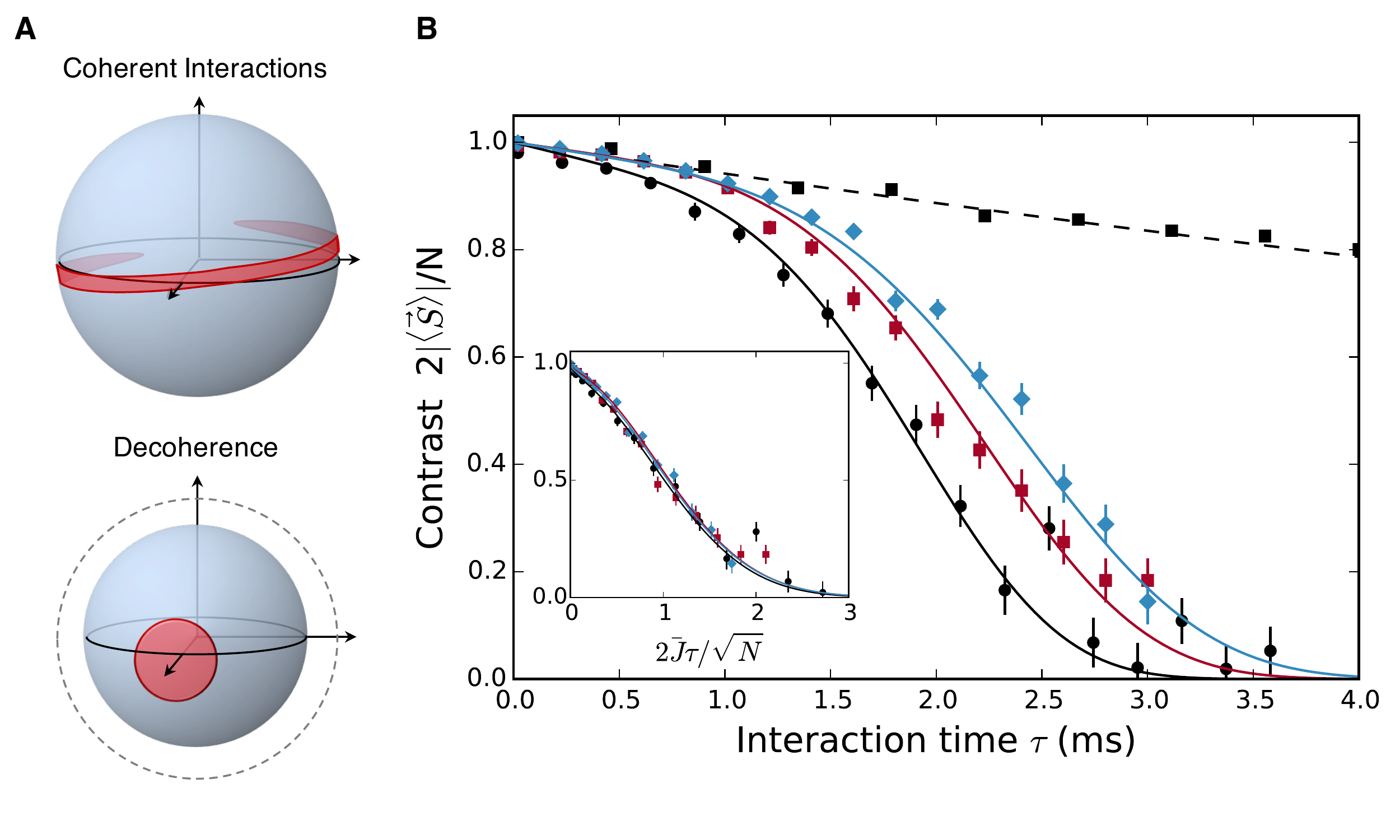}
\caption{{\bf Depolarization of the collective spin from spin-spin interactions and decoherence.} {\bf{(A)}} A diagram of the quasi-probability distribution of the collective spin state on a Bloch sphere, illustrating (top) an over-squeezed state generated by the Ising interaction with no decoherence and (bottom) a loss of contrast only from decoherence, effectively shrinking the Bloch sphere. {\bf{(B)}} Contrast versus interaction time for $N=$ 21, 58, and 144 ions indicated by circles, squares, and diamonds, respectively. The error bars show one standard deviation of the mean, and the solid lines are predictions with no free parameters. The contrast decay from decoherence due to spontaneous emission is measured in the absence of spin-spin coupling (black squares with the dashed line showing an exponential fit). Note that at each $\tau$, the detuning $\delta$ is adjusted to eliminate spin-motion coupling at the end of the experiment, resulting in a different $\Jbar{} \propto1/\delta$ for each point. (Inset): The data collapse to a common curve with proper rescaling, indicating the depolarization is dominated by coherent spin-spin interactions.
	}
\label{fig:depolarization}
\end{figure}

\label{sec:expt}
The depolarization of the collective spin length \vecS{}, or contrast, due to the Ising interaction is a canonical example of non-equilibrium quantum dynamics\cite{Kastner2011,Foss-Feig2013,Hazzard2014}. 
Quantum correlations reduce the contrast and cause the collective spin state to wrap around the Bloch sphere that represents the state space (Fig. \ref{fig:depolarization}A).
However, the contrast also decreases from decoherence, which destroys correlations, effectively shrinking the Bloch sphere.
Our calculation accounts for both effects, and for homogenous Ising interactions $J_{i,j} = \Jbar{}$, the contrast is approximately\cite{SM_sci} given by 
\begin{equation}
	\vecS{} = e^{-\Gamma\tau} \frac{N}{2} \left[ \cos{\left(\frac{2\Jbar{}}{N}\tau \right)} \right]^{N-1}.
	\label{eqn:depol}
\end{equation}
\noindent Here $\tau$ is the total ODF interaction time (Fig. \ref{fig:expt}c) and $\Gamma$ is the total single-particle decoherence rate\cite{SM_sci} due to spontaneous emission from the ODF lasers.

We show the depolarization dynamics of \vecS{} in our experiment in Fig. \ref{fig:depolarization}B, distinguishing effects of coherent interactions from decoherence.
We determine \vecS{} from measurements of $\langle \hat{S}_x \rangle$, performing independent experiments to confirm that $\langle \hat{S}_y \rangle = \langle \hat{S}_z \rangle = 0$ after evolution under $\hat{H}_I$. 
To distinguish the depolarization due to decoherence associated with the ODF lasers alone, we perform experiments at $\delta = +2\pi\times 50$ kHz, effectively eliminating the Ising coupling while leaving the spontaneous emission rate unchanged. 
The dashed line in Fig. \ref{fig:depolarization}B is a fit to the observed exponential decay, measuring $\Gamma$ in our system\cite{SM_sci}.
The significantly faster contrast decay for $\mu$ tuned near $\omega_z$ is in good agreement with Eq. (\ref{eqn:depol}) for a range of system sizes. 
For these data, $\delta = 4\pi/\tau$, ensuring spin-motion decoupling of the COM mode at the end of the experiment\cite{Sawyer2012}.
The collapse of the data to a single curve when plotted as a function of $2 \Jbar{}\tau/\sqrt{N}$, shown in the inset to Fig. \ref{fig:depolarization}B, provides strong evidence that the depolarization is primarily the result of spin-spin interactions. 
However, depolarization dynamics alone are not enough to prove that entanglement exists in the ensemble.

\begin{figure}
\includegraphics[width=5.5in]{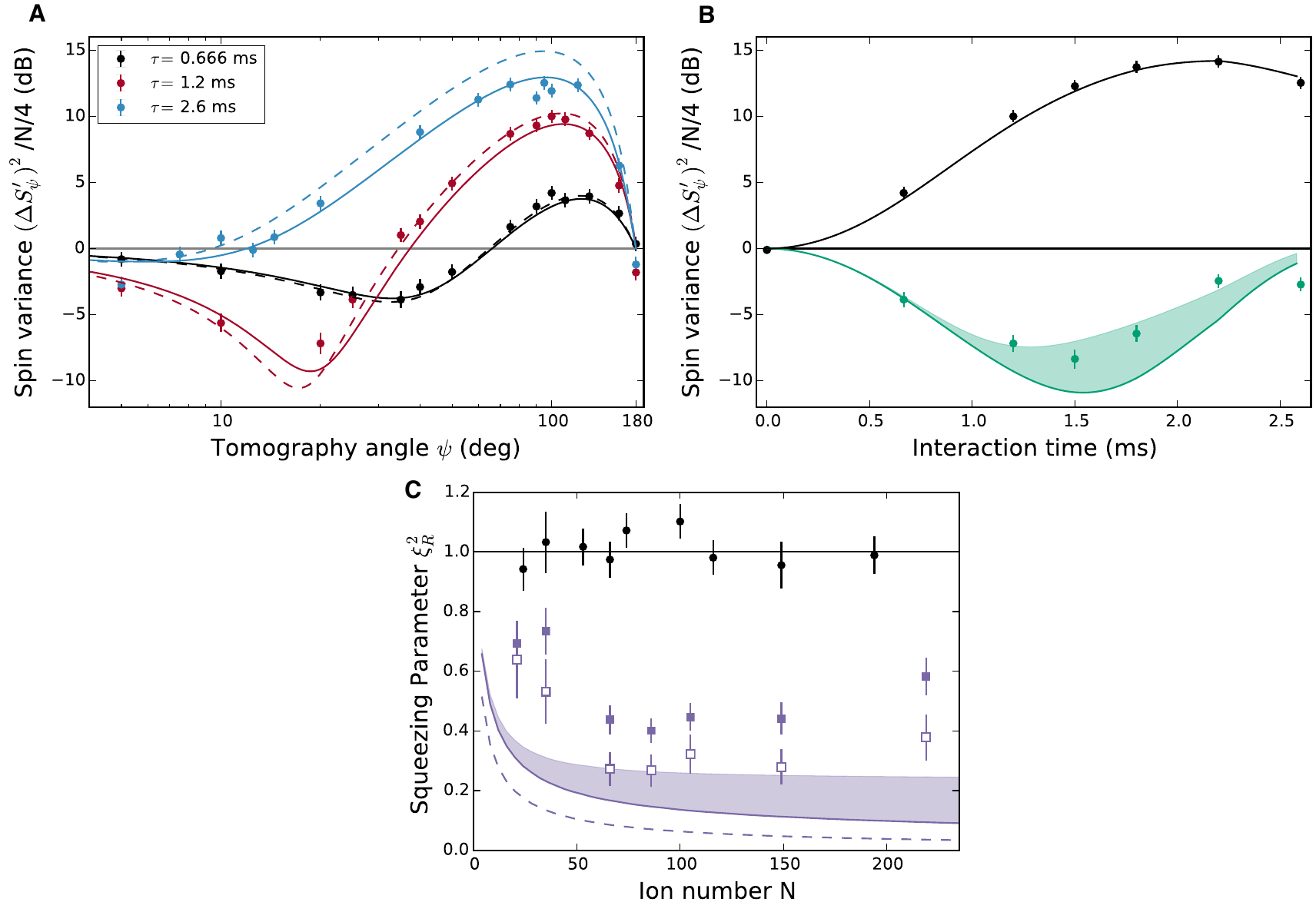}
\caption{{\bf Spin variance and entanglement.} {\bf (A)} Spin variance as a function of tomography angle $\psi$ for $N=$ 86 $\pm$ 2. The variance is calculated from 200 trials. The solid lines are a prediction, with no free parameters, assuming homogenous Ising interactions and including decoherence from spontaneous emission. The dashed lines are theoretical predictions with the same interaction parameters but no decoherence. {\bf (B)} The explicit time dependence of the spin variance for the ensemble in (A). The data for the squeezed (green points) and anti-squeezed (black points) quadratures are shown with theory predictions (solid lines), including decoherence. Since our measurement of $(\Delta S'_\psi)^2$ has significant granularity, we visualize the the effect of finite sampling of $\psi$ using the green shaded region bounded by $\textrm{max} [(\Delta S'_\psi(\psi_m\pm5^\circ))^2]$, where $\psi_m$ corresponds to the angle that minimizes $(\Delta S'_\psi)^2$. The $\pm$5$^\circ$ uncertainty does not have a visible effect in the anti-squeezing component on this plot. {\bf (C)} Ramsey squeezing parameter measured for different ensemble sizes $N$. The black points show data for the initial unentangled spin state. The solid purple squares show the lowest directly measured $\xi_R^2$ with no corrections or subtractions of any detection noise for evaluation of the entanglement witness. The open squares show $\xi_R^2$ inferred by subtracting photon shot noise. The dashed line is the predicted optimal $\xi_R^2$ from coherent Ising interactions with no decoherence, and the solid line shows the limit including spontaneous emission assuming $\Gamma/\Jbar{}=0.05$, which is typical for our system. The shaded purple region accounts for finite sampling of $\psi$ as in (B). All error bars indicate one standard error.}
\label{fig:spin_noise}
\end{figure} 

To verify entanglement, we use the Ramsey squeezing parameter $\xi_R^2$, which only requires measuring the variance of collective observables, instead of full state tomography. 
The Ramsey squeezing parameter is 
\begin{equation}
	\xi_R^2 = N\frac{\textrm{min}_\psi[(\Delta S_\psi)^2]}{\vecS{}^2}\,,
	\label{eqn:xi_r}
\end{equation}
where $\hat{S}_\psi = \frac{1}{2} \sum_i^N\cos(\psi)\hat{\sigma}^z_i+\sin(\psi)\hat{\sigma}^y_i$ and $\textrm{min}_\psi[ \, ]$ indicates taking the minimum as a function of $\psi$.
For an unentangled spin state, polarized along the $x$-axis, $\vecS{}=N/2$ and the spin noise is set by Heisenberg uncertainty relations to $(\Delta S_y)^2  = (\Delta S_z)^2 = N/4$, so $\xi_R^2=1$.
This quantum noise limits the signal-to-noise ratio for a wide range of quantum sensors based on ensembles of independent quantum objects\cite{Wineland1992}. 
Non-classical correlations can redistribute quantum noise between two orthogonal quadratures of the collective spin, squeezing the noise such that $(\Delta S_\psi)^2 < N/4$ and $\xi_R^2<1$. 
These squeezed states are entangled\cite{Sørensen2001c}, and furthermore, $\xi_R^2$ is sufficient to quantify the usefulness of the entanglement as a resource for precise sensing.
As a result, the generation of spin-squeezed states is widely studied \cite{Wineland1992, Shalm2009,Leroux2010c,Wasilewski2010,Hamley2012a,Ockeloen2013,Bohnet2014a,Strobel2014,Behbood2014}. 	

At short times, the non-equilibrium spin dynamics due to the Ising interaction can produce spin-squeezed states\cite{Kitagawa1993,Wineland1992,Leroux2010c,Foss-Feig2013}.
Figures \ref{fig:spin_noise}A and \ref{fig:spin_noise}B show the measured time evolution of the spin variance $(\Delta S'_\psi)^2$ of 86 ions, normalized to the spin variance of the initial, unentangled state.
We compare the data to an analytic model\cite{Foss-Feig2013} that assumes homogenous Ising interactions and fully accounts for both elastic and spin-changing spontaneous emission.
The data clearly show the development of squeezed and anti-squeezed quadratures, and deviations from perfectly coherent Ising dynamics are well described by the effects of spontaneous emission alone.
Similar data for different $N$ are shown in \cite{SM_sci}.

Using measurements of the directly observed spin variance $(\Delta S_\psi)^2$ and contrast $\vecS{}$, we obtain $\xi_R^2$ for a range of $\tau$.
For a given ensemble size $N$, we plot the minimum observed $\xi_R^2$, shown in Fig. \ref{fig:spin_noise}c, where we see that the entanglement witness $\xi_R^2<1$ is satisfied for seven independent datasets with $N$ ranging from 21 to 219. 
We also show $\xi_R^2$ measured for the initial state, confirming our calibration of $N$.
For comparison, Fig. \ref{fig:spin_noise}c shows the absolute minimum $\xi_R^2$ predicted for coherent Ising interactions. 
The majority of the observed discrepancy for ensembles ranging from 60 to 150 ions is accounted for by photon shot noise, spontaneous emission, and the finite sampling of $\tau$ and $\psi$.
For other ion numbers\cite{SM_sci}, we still observe good agreement in the anti-squeezed spin variance, but the minimum spin variance and $\xi_R^2$ deviate further from the prediction.
We attribute the deviation to technical noise sources, described in \cite{SM_sci}.

The Ramsey squeezing parameter is an effective entanglement witness at short times when quantum noise is approximately Gaussian.
At longer times, the growth of spin correlations causes both the depolarization seen in Fig. \ref{fig:depolarization} and the increase in $\textrm{min}_\psi[(\Delta S_\psi)^2]$, due to the appearance of non-Gaussian quantum noise in the collective spin.
Both effects cause $\xi_R^2$ to increase above 1, which we call an over-squeezed state. 
Over-squeezed states can be entangled\cite{Pezze2009}, however, $\xi_R^2$ can also increase simply because of decoherence.

\begin{figure}
\includegraphics[width=4.3in]{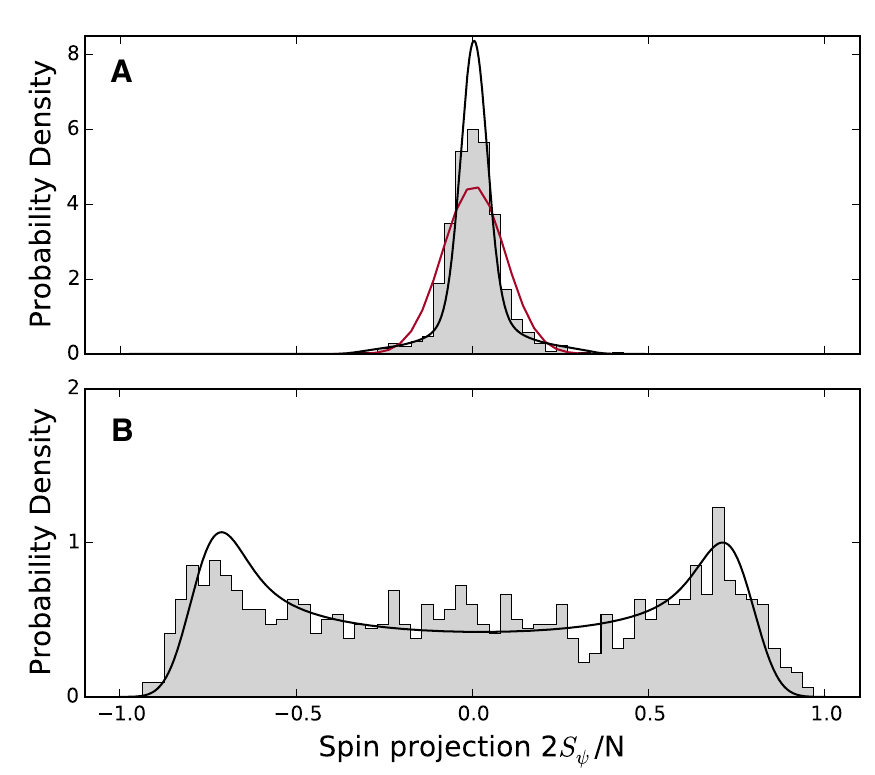}
\caption{{\bf Full counting statistics of a non-Gaussian spin state.} Histograms showing the {\bf{(A)}} squeezed and {\bf{(B)}} anti-squeezed quadratures of the collective spin of $N=$127$\pm$4 ions, corresponding to $\psi =$ 5.4$^\circ$ and $\psi =$ 92$^\circ$, respectively. Here $\tau = $ 3 ms.  The integral of the histogram is normalized to unity. The red line is the Gaussian distribution of the initial, unsqueezed state. The solid black line is the probability density predicted from numerical calculations\cite{SM_sci}, assuming homogenous interactions and including decoherence from spontaneous emission and magnetic field fluctuations. We account for photon shot noise by convolving the theoretical probability density with a Gaussian distribution with a variance $(\Delta S_{psn})^2/(N/4)$.}   
\label{fig:Hist}
\end{figure} 

In order to observe signatures of quantum correlations at longer interaction times, in Fig. \ref{fig:Hist} we show the histogram of the measurements of $\langle \hat{S}_{\psi}\rangle$ for an over-squeezed state of 127 ions after an interaction time of $\tau = $ 3 ms.
For times well beyond the optimum squeezing time, we see a clear non-Gaussian distribution for the anti-squeezed quadrature. The distribution for $\psi=5.4^\circ$ also contains non-Gaussian characteristics in the tails away from the narrow central feature. 
We compare the data to a theoretical model of the full counting statistics and observe good agreement.
Even though $\xi_R^2=26$, the theoretically predicted state is entangled, which we verify using an entanglement witness based on the Fisher information $F$. 
The Quantum Fisher information has been used as an entanglement witness in other trapped-ion simulators\cite{Smith2015}.
We bound the Fisher information using the approach in Ref.\cite{Strobel2014} and find $F/N>2.1$, which satisfies the inequality of the entanglement witness $F/N>1$ \cite{Pezze2009}.
Photon shot noise in our measurement limits our capability to directly witness the entanglement experimentally\cite{SM_sci}, but the good agreement with theory indicates that the state of the ensemble is consistent with an entangled, over-squeezed state.
The full counting statistics are only efficiently computable for homogenous couplings, a good approximation for the small detunings $\delta$ considered here.
For future work with inhomogeneous Ising coupling, obtaining the full counting statistics theoretically will likely be intractable for more than 20 to 30 spins. 

\label{sec:conclusion}
In conclusion, we have verified numerous hallmark signatures of quantum dynamics in ensembles of hundreds of trapped ions, demonstrating spin-squeezing and showing evidence of over-squeezed states, where the magnitude of the collective spin $\vecS{}$ is near zero and the full counting statistics are non-Gaussian.
The techniques presented here are applicable to precision sensors using trapped ions, where the number of ions is limited by systematic errors arising from ion motion\cite{Arnold2015}, and could be useful for quantum-enhanced metrology with non-Gaussian spin states \cite{Strobel2014,Lucke2014,Haas2014,McConnell2015}.
These results benchmark controlled quantum evolution in a 2D platform with more than 200 spins, establishing a foundation for future experiments studying the full transverse-field Ising model in regimes inaccessible to classical computation.
With the implementation of single-spin readout, the simulator could provide unique opportunities to study the dynamics of spin correlations in 2D systems, such as Lieb-Robinson bounds\cite{Richerme2014} and many-body localization in the presence of disorder\cite{Binder1986,Nandkishore2015}.


\paragraph*{Acknowledgements:} The authors acknowledge helpful discussions with David Hume, James Thompson, Adam Keith, Johannes Schachenmayer, Arghavan Safavi-Naini, Martin Gaerttner, and Michael Kastner. All authors acknowledge financial support from NIST. A.M.R. acknowledges support from NSF-PHY 1521080, JILA-NSF-PFC-1125844, ARO, MURI-AFOSR and AFOSR. J.G.B., M.F.F., and M.L.W. acknowledge financial support from the National Research Council Research Associateship Award at NIST. This manuscript is the contribution of NIST and is not subject to US copyright. Correspondence should be addressed to J.G.B. (justin.bohnet@nist.gov) or J.J.B. (jjb@nist.gov).

\include{./SM/SM_penning_squeezed}

\end{document}

%% file: SM/SM_penning_squeezed.tex
\clearpage


\renewcommand\thefigure{S\arabic{figure}}    
\setcounter{figure}{0} 

\renewcommand\theequation{S\arabic{equation}}    
\setcounter{equation}{0} 

\singlespacing

\section*{Supplementary Material}

\subsubsection*{$^{9}$Be$^{+}$ Spin-1/2, Control, and Detection}

\begin{figure}
\includegraphics[scale=1.2]{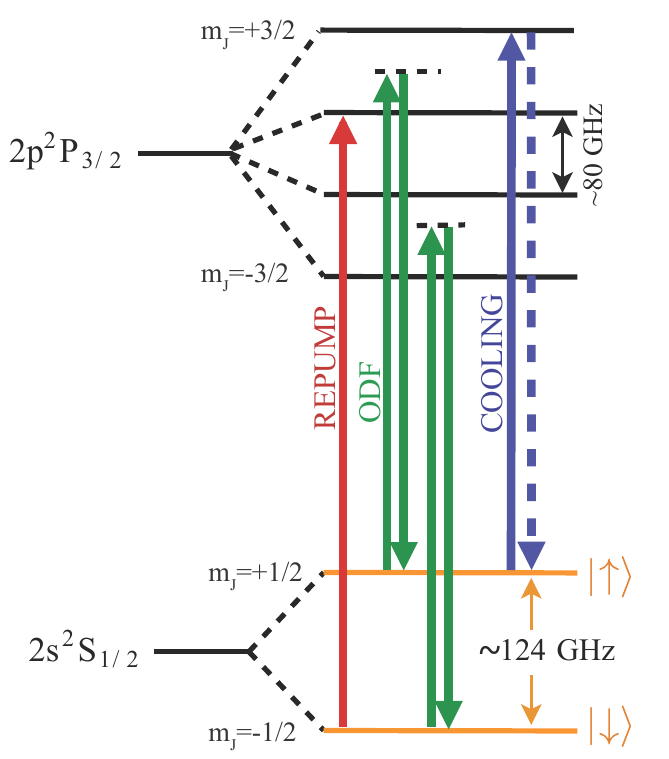}
\caption{Relevant energy levels of $^{9}$Be$^{+}$ at $B_{0}=4.46$~T (not
drawn to scale). We only show $m_{I}=+\frac{3}{2}$ levels which are
prepared experimentally through optical pumping. The $^{2}S_{1/2}-{}^{2}P_{3/2}$
transition wavelength is $313$~nm. A resonant laser beam provides
Doppler laser cooling and state discrimination, a second repumps $\left|\downarrow\right\rangle $
to the $\left|\uparrow\right\rangle $. The ODF interaction is due
to a pair of beams (derived from the same laser) with relative detuning
$\mu$. The qubit splitting $\Omega_{0}\sim2\pi\times124$~GHz. A
low phase noise microwave source at $\Omega_{0}$ provides full global
control over spins. }
\label{figLevelDiagram}
\end{figure}

Reference \cite{Biercuk2009} and the supplementary material to \cite{Britton2012}
give detailed descriptions of our spin initialization, control, and
measurement capabilities with planar ion arrays in Penning traps.
We briefly summarize some of that discussion, emphasizing aspects
relevant for the measurements reported here. Figure \ref{figLevelDiagram}
shows the relevant $^{9}$Be$^{+}$ energy levels. We use the valence
electron spin states parallel $\left|\uparrow\right\rangle =\left|m_{J}=+\frac{1}{2}\right\rangle $
and anti-parallel $\left|\downarrow\right\rangle =\left|m_{J}=-\frac{1}{2}\right\rangle $
to the applied magnetic field of the Penning trap as the spin-$1/2$
or qubit. In the 4.46 T magnetic field of the trap these levels are
split by approximately $\Omega_{0}$=$2\pi\times124$ GHz. The $^{9}$Be$^{+}$
nucleus has spin $I=3/2$. We optically pump the nuclear spin
to the $m_{I}=+3/2$ level \cite{Itano1981}, where it remains throughout the duration
of an experiment. The ions are Doppler laser-cooled to a temperature
$\sim$0.5 mK by two 313 nm laser beams tuned approximately 10 MHz below
the $\left|\uparrow\right\rangle \rightarrow\left|^{2}P_{3/2}\: m_{J}=+3/2\right\rangle $
cycling transition and directed parallel and perpendicular to
the magnetic field \cite{Jensen2004}. Spins in the $\left|\downarrow\right\rangle $
state are efficiently optically pumped to the $\left|\uparrow\right\rangle $
state with a laser tuned to the $\left|\downarrow\right\rangle \rightarrow\left|^{2}P_{3/2}\: m_{J}=+1/2\right\rangle $
transition. A typical experimental cycle starts with $\sim3$ ms
of combined Doppler laser cooling and repumping. We estimate
the fidelity of the $\left|\uparrow\right\rangle $
state preparation should be very high ($\gg99$\%).

Low-phase noise microwave radiation from a 124 GHz source described previously
in the supplementary material to \cite{Britton2012} is used to globally rotate the spins,
and provides an effective transverse magnetic field in the rotating
frame of the qubit. The length of the time interval required to
drive $\left|\uparrow\right\rangle$ to $\left|\downarrow\right\rangle$ ($\pi$-pulse) was
$\sim60 \mu$s. The fidelity of a $\pi$-pulse was measured to
be greater than 99.9\% in a random benchmarking experiment \cite{Biercuk2009}. The spin-echo
coherence duration ($T_2$) was measured to be $\sim10$ ms with the magnet sitting
on the floor, and greater than $50$ ms with the magnet vibrationally isolated.  All
measurements were done with the magnet sitting on the floor.

At the end of an experimental sequence we turn on the Doppler cooling
laser and make a projective measurement of the ion spin state through
state-dependent resonance fluorescence. With the Doppler cooling laser
on, an ion in the $\left|\uparrow\right\rangle $ state scatters photons
while an ion in $\left|\downarrow\right\rangle $ is dark. Specifically
we detected, with f/5 light collection and a photomultiplier tube
(PMT), the resonance fluorescence from all the ions in a direction
perpendicular to the magnetic field (the side-view). The PMT counts
are integrated over time periods of typically 1 ms for measuring averages, or
15 ms for measuring standard deviation (to reduce the impact of photon
shot noise). The photon detection rate varied between 1 and 2 photons
per ion per ms. The variation is due to day-to-day variation in the
cooling beam intensities and positions, usually changed to optimize
crystal stability. Once data collection started, the photon detection
rate was held constant by fixing the Doppler cooling beams' positions
and stabilizing their intensities. The integrated photon count is
converted to state population measurement using a frequently repeated
calibration of the counts for all the spins in $\left|\uparrow\right\rangle $
and for all spins in $\left|\downarrow\right\rangle $.

Images of the ions in the rotating frame of the crystal were obtained by using
an imaging PMT (maximum processing capability $\lesssim$ 100 kHz)
to record ($x$,$y$,$t$) for each photon \cite{Mitchell2001}. These images were used to count
the number of $^{9}$Be$^{+}$ ions in the crystal, and were recorded
either before or after long sequences of measurements. The number
of $^{9}$Be$^{+}$ ions slowly decreases due to the formation of
BeH$^{+}$ through collisions with residual H$_{2}$ in the room-temperature
vacuum system. This slow change in the $^{9}$Be$^{+}$ ion number
was tracked by monitoring the resulting slow change in the global
fluorescence. Reversing the hydride-ion formation through photodissociation of BeH$^{+}$
has been demonstrated \cite{Sawyer2015} and can be implemented in the future with a redesign
of the vacuum envelope.

We employ a recently designed and fabricated Penning trap consisting
of a stack of cylindrical electrodes (see sketch in Fig. 1 of the
main text) to generate an electrostatic potential $q\phi_{trap}(\rho,z)\simeq\frac{1}{2}m\omega_{z}^{2}\left(z^{2}-\rho^{2}/2\right)$
near the center of the trap, where $z$ and $\rho$ are cylindrical
coordinates. With potentials of up to 2 kV we obtain $\omega_{z}\simeq 2\pi\times1.57\,\textrm{MHz}$
while nulling the lowest order anharmonic ($C_{4}$) term. The direction
of the magnetic field of the trap is aligned with the symmetry axis (the
$z$-axis) of the electrodes to better than $0.01^{\circ}$. The rotation
frequency $\omega_r$ of the ion array determines the strength of the radial confinement
of the ion crystal, and is precisely controlled by a rotating quadrupole
potential \cite{Huang1998b,Dubin1999,Hasegawa2005}. The middle electrode of the trap incorporates eight azimuthally
segmented rotating wall electrodes at a radius of 1 cm from the center
of the trap. This configuration enables control of smaller arrays
than possible in our previous trap, presumably because stronger rotating
wall potentials are easily generated. For most of the measurements
recorded here, the rotating wall potential is characterized by $q\phi_{2,2}(x_{R},y_{R})=\frac{1}{2}m\omega_{q}^{2}\left(x_{R}^{2}-y_{R}^{2}\right)$
where $\omega_{q}\simeq2\pi\times(28\,\textrm{kHz})$ and $x_{R},y_{R}$
denote coordinates in the rotating frame.

\subsubsection*{Optical-Dipole Force and Lamb-Dicke Confinement}

A spin-dependent optical dipole force is obtained from a moving 1D
optical lattice generated at the intersection of two off-resonant
laser beams. The set-up is identical to that described in \cite{Sawyer2012,Sawyer2014}
and in the supplementary material to \cite{Britton2012}, except the beams
cross with an angle $\theta=20^{\circ}$ ($\pm10^{\circ}$
with respect to the central $z=0$ plane of the trap). The optical
dipole force (ODF) beams are detuned by approximately 20 GHz from
any electric dipole transitions in Fig. \ref{figLevelDiagram} and
produce an AC Stark shift on the $\left|\uparrow\right\rangle $ and
$\left|\downarrow\right\rangle $ states,
\begin{align}
\begin{array}{ccc}
\Delta_{\uparrow,acss} & =  \epsilon_{\uparrow}+\frac{1}{2}U_{\uparrow}\sin\left[\delta\vec{k}\cdot\vec{\hat{r}}-\mu t\right]\\
\Delta_{\downarrow,acss} & =  \epsilon_{\downarrow}+\frac{1}{2}U_{\downarrow}\sin\left[\delta\vec{k}\cdot\vec{\hat{r}}-\mu t\right]
\end{array}\:.\label{eq:state ACSS}
\end{align}
Here $\delta\vec{k}$ and $\mu$ are the wave vector and frequency
difference between the ODF beams. $\left|\delta\vec{k}\right|=2k\sin\left(\theta/2\right)=2\pi/\left(0.90\,\mu\textrm{m}\right)$
for $\theta=20^{o}$ . We adjust the polarization and frequency of the ODF laser beams so 
that $\epsilon_{\uparrow}=\epsilon_{\downarrow}$ and $U_{\uparrow}=-U_{\downarrow}\equiv U$,
producing a spin-dependent ODF potential,
\begin{equation}
\hat{H}_{ODF}=U\sum_{i}\sin\left[\delta\vec{k}\cdot\vec{\hat{r}}_{i}-\mu t\right]\hat{\sigma}_{i}^{z}\,.\label{eq:full ODF}
\end{equation}
To minimize the variation of the phase of the 1D optical lattice across
the ion array we align $\delta\vec{k}\parallel\hat{z}$.  We do this by
minimizing decoherence of the spins with $\hat{H}_{ODF}$ applied
and $\mu$ tuned to a harmonic of the rotation frequency, $\mu=n\omega_{r}$.
By comparing the $n=2$ and $n=1$ decoherence signals, we estimate
a misalignment error $\left|\Delta\theta_{err}\right|\lesssim 0.01^{o}$.
For $N=200$ (largest numbers used for data collection), the array
radius is $R_{0}\simeq125\,\mu\textrm{m}$, and the phase difference between the center and edge of the array is $\left|\delta\vec{k}\right|R_{0}\tan\left(\Delta\theta_{err}\right)\lesssim 0.15$ radians $\simeq 8.5^{\circ}$.

With $\delta\vec{k}\parallel\hat{z}$, Eqn. \ref{eq:full ODF} only
involves the axial coordinates of the ions, \\$\hat{H}_{ODF}=U\sum_{i}\sin\left[\delta k\,\hat{z}_{i}-\mu t \right]\hat{\sigma}_{i}^{z}$, where $\delta k=\left|\delta\vec{k}\right|$.
In the Lamb-Dicke confinement limit ($\delta k\,z_{rms,i}\ll1$,
$z_{rms,i}\equiv\sqrt{\left\langle \hat{z}_{i}^{2}\right\rangle }$
is the root mean square (rms) axial extent of the wave function of
ion $i$), this reduces to $\hat{H}_{ODF}\simeq F_{0}\cos\left(\mu t\right)\sum_{i}\hat{z}_{i}\hat{\sigma}_{i}^{z}$
(assuming $U/\mu\ll 1)$ where $F_{0}\equiv U\cdot\delta k$. The spin-dependent
ODF, $F_{0}$, is reduced outside of the Lamb-Dicke confinement limit
by the Debye-Waller factor $DWF_{i}\equiv\exp\left(-\frac{1}{2}\delta k^{2}z_{rms,i}^{2}\right)$
\cite{wineland1998bible}. The Ising
pair-wise coupling strengths are reduced by the square of the Debye-Waller
factor. The measured Ising coupling strengths, determined from mean-field
spin precession measurements, were less than the calculated coupling
strengths (i.e. $U\cdot\delta k$) by $10\%$ to $25\%$, in rough agreement with the Debye-Waller
factors estimated below. Typical values for this work are $U\simeq\hbar 2\pi\times(6.5\,\textrm{kHz})$
resulting in $F_0=30 \,\textrm{yN}$.  The calculated coupling strengths are based
on well known atomic physics parameters for $^{9}$Be$^{+}$, and
calibration of the ODF laser intensity through AC Stark shift measurements
for different polarizations of the ODF beams.

Measurements indicate the temperature of the axial drumhead modes
is close to the Doppler cooling limit of $\sim0.5$ mK \cite{Sawyer2012,Sawyer2014}.
Neglecting the Coulomb interaction between the ions, we estimate $z_{rms,i}\simeq\sqrt{\frac{\hbar}{2m\omega_{z}}\left(2\bar{n}+1\right)}\simeq71$
nm, $\delta k\cdot z_{rms,i}=0.50$, and $DWF_{i}\simeq0.88$. An
improved estimate of $z_{rms,i}$ is obtained by summing the contributions
from all of the transverse modes $m$
\begin{equation}
z_{rms,i}=\left(\sum_{m}\left(b_{i,m}\right)^{2}\frac{\hbar}{2m\omega_{m}}(2\bar{n}_{m}+1)\right)^{1/2}\,\label{eq:axial rms}
\end{equation}
where $\bar{n}_{m}\simeq k_{B}T/\hbar\omega_{m}$ and $b_{i,m}$ is 
the amplitude of the $m^{\mathrm{th}}$ normal mode at site $i$. With $N=127$,
$\omega_{r}=2\pi\times180$ kHz, $\omega_{z}=2\pi\times1.575$ MHz, and 
$T=0.5$ mK (typical parameters used in this work), $z_{rms,i}\simeq77$
nm, $\delta k\cdot z_{rms,i}=0.54$, $DWF_{i}\simeq0.86$ in the
center of the array, changing to $z_{rms,i}\simeq72$, $\delta k\cdot z_{rms,i}=0.50$,
$DWF_{i}\simeq0.88$ at the radial edge of the array.

In addition to reducing the average strength of the spin-dependent
ODF, a non-zero Lamb-Dicke confinement parameter gives rise to fluctuations
in the spin-dependent ODF from one realization of the experiment to
the next \cite{wineland1998bible}.
These fluctuations can produce fluctuations in the induced spin-spin
interactions. This appears to be a challenging problem to accurately
model for a many-ion array, but large numbers of ions will tend to
average out the effects of thermal motional fluctuations. For our
work where $\mu-\omega_{z}$ is small compared to $\mu-\omega_{m}$
for any non-COM mode $m$, thermal
motional fluctuations give rise to fluctuations in the single-axis twisting
strength produced by the spin-dependent coupling to the COM mode.
We estimate fractional fluctuations in the single-axis twisting strength
to be less than 3\% for $N=100$ and $ $$T=0.5$ mK. 
Sub-Doppler cooling the axial drumhead modes can reduce $z_{rms,i}$.

\subsection*{Spin Variance Measurements with Different Ion Numbers}

\begin{figure}
\includegraphics[scale=0.90]{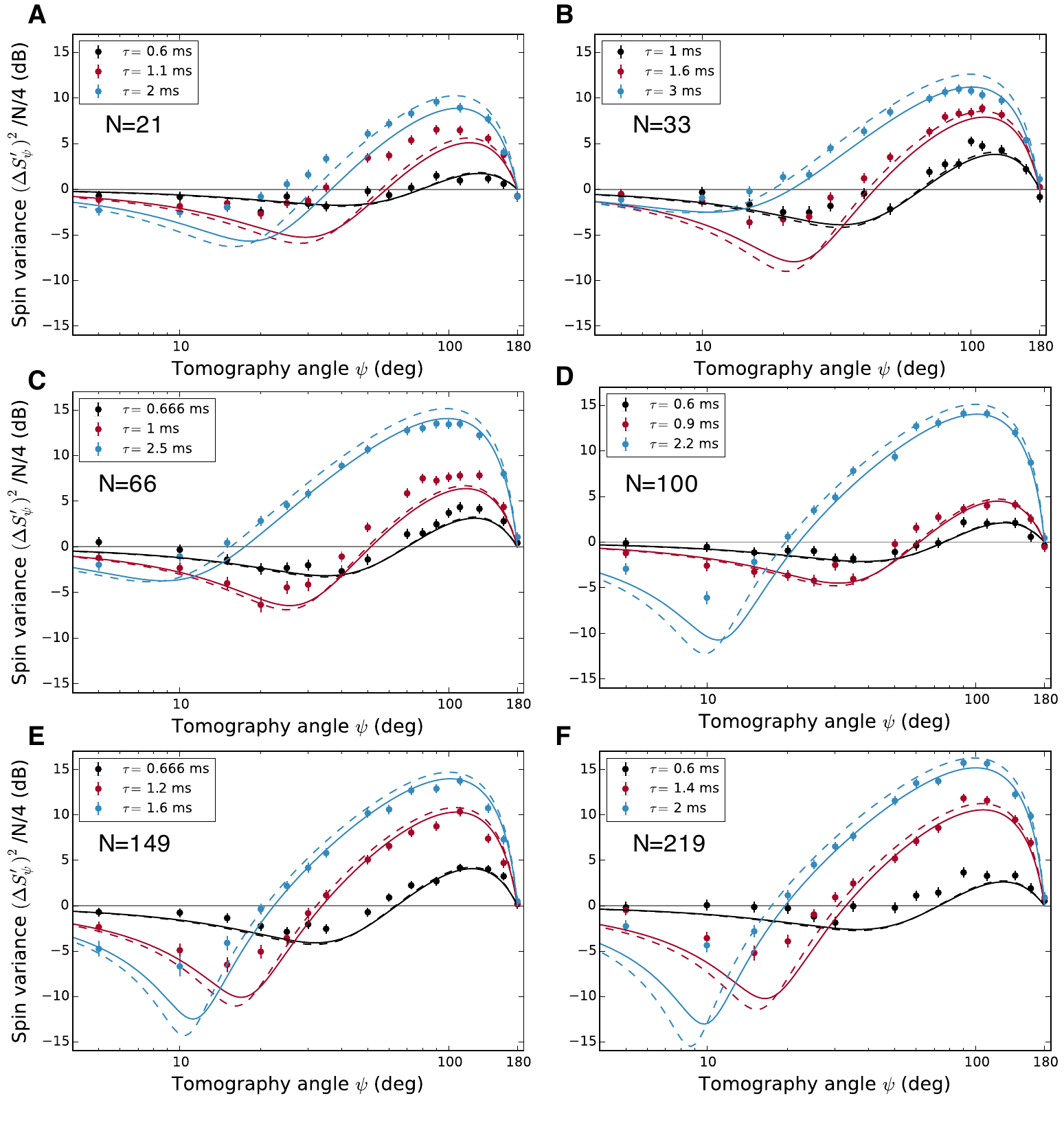}
\caption{Spin variance as a function of tomography angle $\psi$ for different ion numbers $N$, calculated from $N_{\mathrm{trials}}=200$ trials.  The error bars are one standard error on the variance. The solid lines are a prediction, with no free parameters, assuming homogenous Ising interactions and including decoherence from spontaneous emission. The dashed lines are a theoretical prediction with the same interaction parameters but no decoherence. }
\label{fig:ExtraSpinNoise}
\end{figure}

Figure~\ref{fig:ExtraSpinNoise} shows spin variance measurements, analogous to Fig 3A of the main text, with different numbers of ions.
The uncertainty, one standard error $\sigma_S$, in the measured variance $(\Delta S_\psi)^2$ is calculated as $\sigma_{S} = (\Delta S_\psi)^2 \sqrt{2/N_{\mathrm{trials}}}$ where $N_{\mathrm{trials}}$ is the number of experimental trials.  
Then in Fig. \ref{fig:ExtraSpinNoise} and Fig. 3 of the main text, one standard error on the normalized variance is determined by following standard error propagation.

\subsection*{Sources of Noise and Decoherence}

For the analysis in the main text, we account for decoherence due to spontaneous light scattering and photon shot noise.
In particular we model the measured variance in the transverse spin $\left(\Delta S_{\psi}\right)^{2}$ as
\begin{equation}
\left(\Delta S_{\psi}\right)^{2}=\left(\Delta S_{\psi}|_{\Gamma}\right)^{2}+\frac{m}{K^{2}}\,.\label{eq:noise model}
\end{equation}
Here $\left(\Delta S_{\psi}|_{\Gamma}\right)^{2}$ denotes the prediction for the transverse spin variance obtained with the engineered Ising interaction in the presence of spontaneous light scattering from the ODF laser beams \cite{Foss-Feig2013}. 
The contribution of photon shot noise to the variance is $(\Delta S_{psn})^2 = m/K^{2}$.
Here $m$ is the mean number of photons collected in a global fluorescence measurement and $K$ is the number of photons collected per ion in the bright state $\left|\uparrow\right\rangle $. 
The angle $\psi$, defined in the main text, denotes the angle along which the transverse spin variance is measured. 
In the main text, $\left(\Delta S_{\psi}\sp{\prime}\right)^{2}\equiv\left(\Delta S_{\psi}\right)^{2}-(\Delta S_{psn})^2 $.
We separately discuss decoherence due to spontaneous emission and photon shot noise.
We also discuss a few potential sources of decoherence that may be contributing to the increase in the variance of the squeezed spin quadrature observed with increasing ion number.

\subsubsection*{Spontaneous emission}

The primary source of decoherence in the simulator arises from spontaneous
emission from the off-resonant ODF laser beams. Decoherence due to
spontaneous light scattering from an off-resonant laser beam has been
carefully studied in this system \cite{Uys2010}. The off-diagonal
elements of the density matrix for an individual spin decay exponentially
with rate $\Gamma\equiv\left(\Gamma_{\mathrm{el}}+\Gamma_{\mathrm{Ram}}\right)/2$
where $\Gamma_{\mathrm{el}}$ and $\Gamma_{\mathrm{Ram}}$ are the decoherence rates
for elastic and Raman scattering, respectively. $\Gamma_{\mathrm{Ram}}=\Gamma_{\mathrm{ud}}+\Gamma_{\mathrm{du}}$ 
where $\Gamma_{\mathrm{ud}}$ and $\Gamma_{\mathrm{du}}$ are the rates for spontaneous transitions from $|\uparrow\rangle$ to $|\downarrow\rangle$ and from $|\downarrow\rangle$ to $|\uparrow\rangle$, respectively. Reference \cite{Uys2010}
provides expressions for $\Gamma_{\mathrm{el}}$ and $\Gamma_{\mathrm{Ram}}$
in terms of atomic matrix elements, and laser beam polarizations and
intensities. For our set-up, $\Gamma_{\mathrm{el}}\sim4\Gamma_{\mathrm{Ram}}$. 

We use the ions to measure the individual laser beam intensities (typically
$\sim0.5\,\textrm{W}/\textrm{cm}^{2}$) through measurements of the
AC Stark shift with the polarization rotated parallel to $\hat{z}$
(the magnetic field axis). We directly measure $\Gamma$ by measuring
the exponential decrease in $|\langle\vec{S}\rangle|$ as a function
of the time interval the ODF beam(s) is (are) turned on. We
use a spin-echo sequence similar to that described in \cite{Uys2010}
and illustrated in Fig. 1 of the main text. We observe good
agreement between the calculated and measured decoherence rates with
the application of a single ODF laser beam. A typical single beam decoherence
rate is $\Gamma\simeq21.5\,\textrm{s}^{-1}$. With the 1D optical
lattice and an ODF beat note $\mu$ tuned $\sim50$ to $\sim100$
kHz above the axial COM mode, we observe exponential decay (dashed
line in Fig. 2 of the main text) of the system Bloch vector at a
rate $\sim50\%$ higher than the sum of the rates from
each beam. This excess decoherence rate is observed to be relatively
independent of the ODF beat note $\mu$, and is presently not understood.

For each data set, measurements of the decoherence rate with $(\mu-\omega_{z})\sim 2\pi\times50$
kHz were used, along with the measured Ising interaction strength
and the theory of \cite{Foss-Feig2013}, to generate $\left(\Delta S_{\psi}|_{\Gamma}\right)^{2}$,
displayed by the solid lines in Fig. \ref{fig:ExtraSpinNoise} and Fig. 3 of the main text. We assume
the measured excess decoherence is due to an increase in $\Gamma_{\mathrm{el}}$. More 
details on the theoretical modeling is given in a subsequent section.
The impact of decoherence due to spontaneous light scattering can be decreased by increasing
the angle $\theta$ with which the ODF beams cross.

\subsubsection*{Photon shot noise, classical detection noise}

Photon shot noise contributes to the measured transverse spin variance.
For a global fluorescence measurement where $m$ photons are collected,
the variance in the number of collected photons is $m$. We assume
the same contribution of photon shot noise to the variance of a series
of global fluorescence measurements where the mean number of photons
collected per measurement is $m$. For measuring the variance of a
spin component, typical detection times were 15 ms resulting in at
least $K=15$ photons collected for an ion in the bright state. For
a spin state in the equatorial plane of the Bloch sphere, the mean
number of photons collected is $m=\frac{N}{2}K$, so photon
shot noise will contribute to $\left(\Delta S_{\psi}\right)^{2}$
at the level $\frac{m}{K^{2}}=\frac{N}{4}\frac{2}{K}$. Relative to
projection noise $\frac{N}{4}$, photon shot noise contributes
at the level of $\frac{2}{K}$, or less than $13 \%$ $(-8.8\:\textrm{dB})$ for $K>15$.
For the variance measurements in Fig. 3 of the main text, the photon shot noise is
subtracted from the measured spin variance.  Shot noise was accounted for in the 
counting statistics of Fig.~4 in the main text by convolving the theoretical probability 
distribution with the distribution of shot noise, as discussed in more detail in a later section.
The relative contribution
of photon shot noise can be reduced with longer detection times. We
estimate the detection time interval can be increased by an order of
magnitude before optical pumping between $\left|\downarrow\right\rangle $
and $\left|\uparrow\right\rangle $ is a consideration.

Classical fluctuations in the detection laser power, frequency, and position can
contribute to the measured spin variance in our global fluorescence
detection. We measure this classical detection noise $\sigma_{t}^{2}$ by initializing
all the ions in $\left|\uparrow\right\rangle $ (bright state) and
measuring the variance in the total photon count
$\sigma_{total}^{2}$, and then infer $\sigma_{t}^{2}=\sigma_{total}^{2}-m$.
We measured $\sigma_{t}^{2}$ to be less than $30\%$ of photon 
shot noise (-14 dB below projection noise).
We neglect this small contribution of classical detection noise relative to
photon shot noise in our analysis. Specifically we do not subtract
any classical detection noise.

\subsubsection*{Magnetic field fluctuations}

\begin{figure}
\includegraphics[scale=0.75]{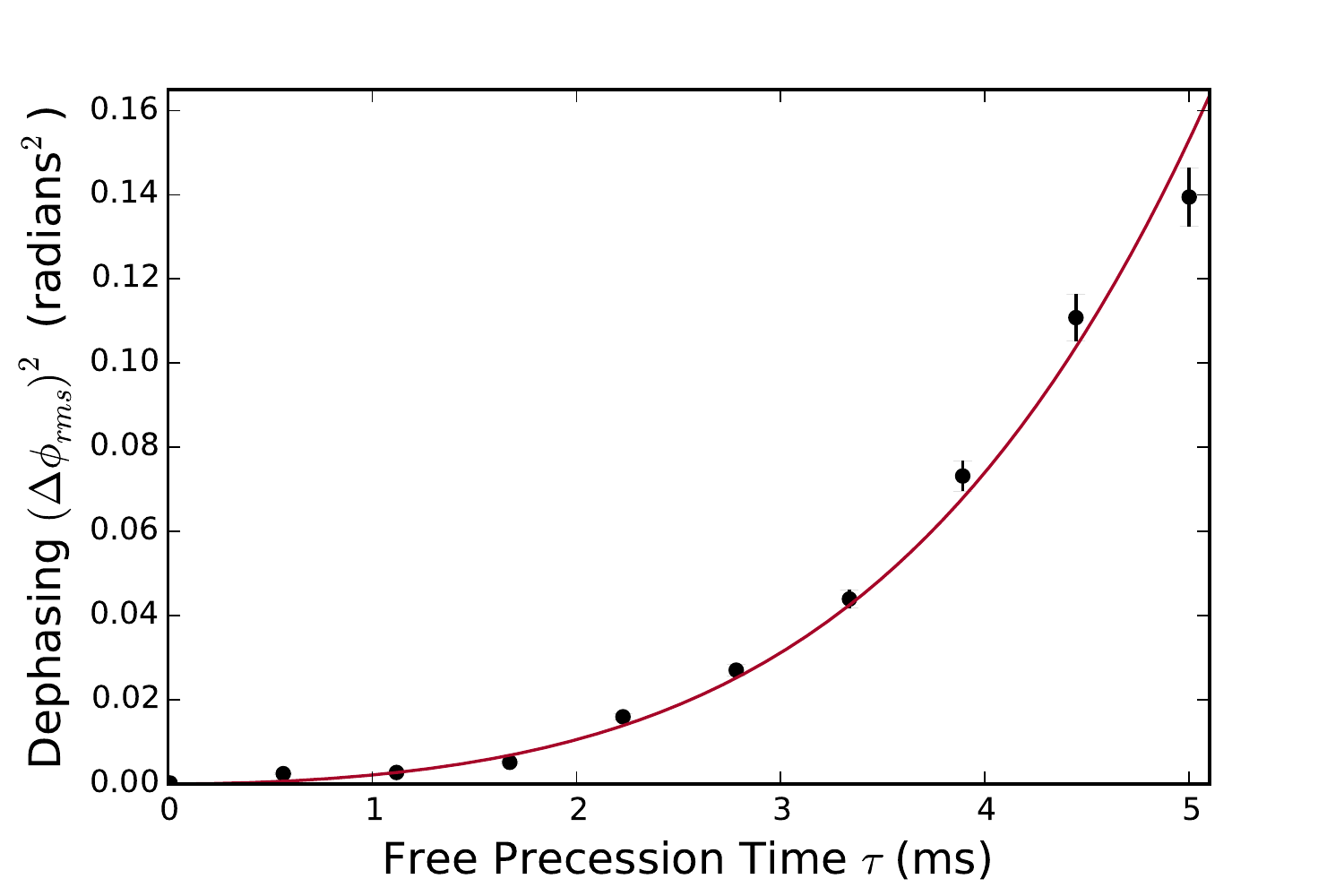}
\caption{Dephasing due to magnetic field fluctuations measured with $N=124(3)$ ions in the absence of the ODF beams.
Plotted is the variance of the dephasing angle $\Delta\phi(\tau)$ determined from 300 trials of a spin echo
experiment measuring the transverse spin noise along $\psi=90^{\circ}$. Photon shot noise was
subtracted; $\tau$ is the sum of the two free precession intervals.  The red line is
a 2-parameter fit $\Delta\phi_{rms}^2(\tau)=(2.4\times10^{-3}/\textrm{ms}^2)\,\tau^2+(1.7\times10^{-4}/\textrm{ms}^4)\,\tau^4$ where $\tau$ is in ms.}
\label{Bfield fluctuations}
\end{figure}

We measure a small amount of dephasing when running the
experiments described in the text in the absence of the ODF lasers.
This is due to fluctuations in the homogeneous magnetic field produced by vibrations
of the superconducting magnet \cite{Biercuk2009,Britton2015}. Without the ODF beams,
homogeneous fluctuations produce a dephasing
proportional to the square of the Bloch vector length, $N^{2}/4$.
We write its contribution to the transverse spin variance
as $\left(N^{2}/4\right)\Delta\phi_{rms}^{2}(\tau)\sin^{2}\left(\psi\right)$
where $\Delta\phi_{rms}^{2}(\tau)$ only depends on the magnetic field noise
spectrum and the length $\tau$ of the experimental sequence \cite{Biercuk2009a,Cywinski2008}.
For spin-echo sequences and magnetic field noise dominated by low frequencies \cite{Britton2015}, we anticipate $\Delta\phi_{rms}^{2}(\tau)\propto\tau^{4}$.
Figure \ref{Bfield fluctuations} shows dephasing measurements obtained without
the ODF beams. Both $\tau^{2}$ and $\tau^{4}$
dependences are observed. By taking care to minimize sources of vibration
in the lab, the measured $\Delta\phi_{rms}^{2}(\tau)$ did not significantly
vary from day to day.

Dephasing is described by the Hamiltonian $B(t)\sum_i \hat{S}^{z}_i$ where $B(t)$
is a stochastic process. This Hamiltonian commutes with the Ising interaction
(and also with the elastic Rayleigh scattering decoherence in the master equation).  The impact of
magnetic-field-induced dephasing can therefore be accurately modeled, and we find
its contribution to be small compared to photon shot noise
for both the variance measurements (Fig. 3, main text) and the histograms (Fig. 4, main text). A more complete discussion of the impact of magnetic field fluctuations is given in a later section.
We note that relative to projection noise $(N/4)$, the contribution of homogeneous dephasing
scales as the length of the Bloch vector $(\propto N)$, becoming more important for larger numbers of ions.
The measured magnetic field noise can be reduced by more than a factor of
5 by vibrationally isolating the magnet.

\subsubsection*{Other potential sources of noise}

We briefly discuss a few other potential sources of dephasing that do
not appear to significantly contribute to the work discussed here, but could become factors, in
particular if photon shot noise is reduced. 

Heating of the axial COM mode during application of the spin-dependent
force is a source of dephasing. Following the discussion in Ref. \cite{Ozeri2007},
we calculate the dephasing $\Delta\phi_{rms}^{2}(\Delta n)$
due to a stochastic increase $\Delta n$ in the COM mode occupation
number during the application of a spin-dependent force $F_{0}$ for
a time $\tau_{s}=2\pi/\delta$ where $\delta=\mu-\omega_{z}$. A spin
echo sequence consists of two such applications, and results in twice
the dephasing (in variance),
\begin{align}
\frac{\Delta\phi_{rms}^{2}(\Delta n)}{\Delta\phi_{proj}^{2}}\simeq\Delta n\cdot\frac{8F_{0}^{2}z_{0}^{2}}{\hbar^{2}\delta^{2}}\,.
\end{align}
Here $\Delta\phi_{proj}^{2}=1/N$ is the angle determined by the projection noise limit and $z_{0}=\sqrt{\hbar/\left(2m\omega_{z}\right)}$.
For close detunings $\delta=2\pi\times1\,\textrm{kHz}$, $\frac{2F_{0}z_{0}}{\hbar\delta}\sim1$,
so $\Delta n\sim1/\textrm{ms}$ can cause dephasing on the order of projection noise. For trapped ions,
the COM mode is typically heated by noisy electric fields.  In this case,
the heating rate scales linearly with $N$ \cite{Sawyer2014}. Measurements
place an upper limit on the COM mode heating rate of $1\,\textrm{(quanta/s)/ion}$.
For $\tau_s=1$ ms and $N$=100 ions, $\Delta n \leq 0.1$.  Because the COM mode
heating rate may scale linearly with $N$, this source of dephasing
will likely become more important as the ion number increases.  We note that photon recoil
from spontaneous light scattering with the ODF beams will produce dephasing
by the mechanism described above.  We estimate its contribution to be small compared
with the $\Delta n=0.1$ estimate.

For small detunings $\delta$ from the COM mode, fluctuations and
drifts in the axial COM mode frequency $\omega_{z}$ can produce spin-motion
entanglement because the decoupling condition $\tau_s=2\pi/\delta$
may no longer be satisfied. Here $\tau_s$ is the duration of a single arm
period of the spin-echo sequence. Spin-motion entanglement produces
dephasing, and we calculate this dephasing with Eqs. (30) and
(32) of Ref. \cite{Sawyer2014}. Let $\delta\tau_s=2\pi+\epsilon$
where $\epsilon$ is a measure of the incomplete full circle due to
error in measuring $\omega_{z}$. We note that a spin echo sequence
suppresses the error due to a non-zero $\epsilon$ (relative to a
 Ramsey sequence \cite{Hayes2012}), and calculate
a dephasing,
\begin{align}
\frac{\Delta\phi_{rms}^{2}(\epsilon)}{\Delta\phi_{proj}^{2}}\simeq\frac{F_{0}^{2}z_{0}^{2}}{\hbar^{2}\delta^{2}}\epsilon^{2}\left(\epsilon+\delta t_{\pi}\right)^{2}\left(2\bar{n}+1\right)\,.
\end{align}
Here $t_{\pi}\simeq60\:\mu\textrm{s}$ is the duration of the $\pi$-pulse
in the spin echo sequence. For $\delta=2\pi\times(1\,\textrm{kHz)}$,
$F_{0}\simeq30\:\textrm{yN}$, $\omega_{z}=2\pi\times(1.6\:\textrm{MHz)}$,
$T_{COM}=1.0\:\textrm{mK}\,(\bar{n}_{COM}\simeq12)$, we estimate
$\Delta\phi_{rms}^{2}(\epsilon)/\Delta\phi_{proj}^{2}<1$ requires $\epsilon<0.5$.
This places an upper limit on the uncertainty of the axial COM mode
frequency of $\Delta\omega_{z}<2\pi\times 80\:\textrm{Hz}$. During data
collection we checked for a shift in the COM mode frequency every 2 s, and
used this information to update $\mu$ to fix $\delta=2\pi/\tau_s$.

\begin{figure}[t]
\begin{centering}
\includegraphics[width=0.65\columnwidth]{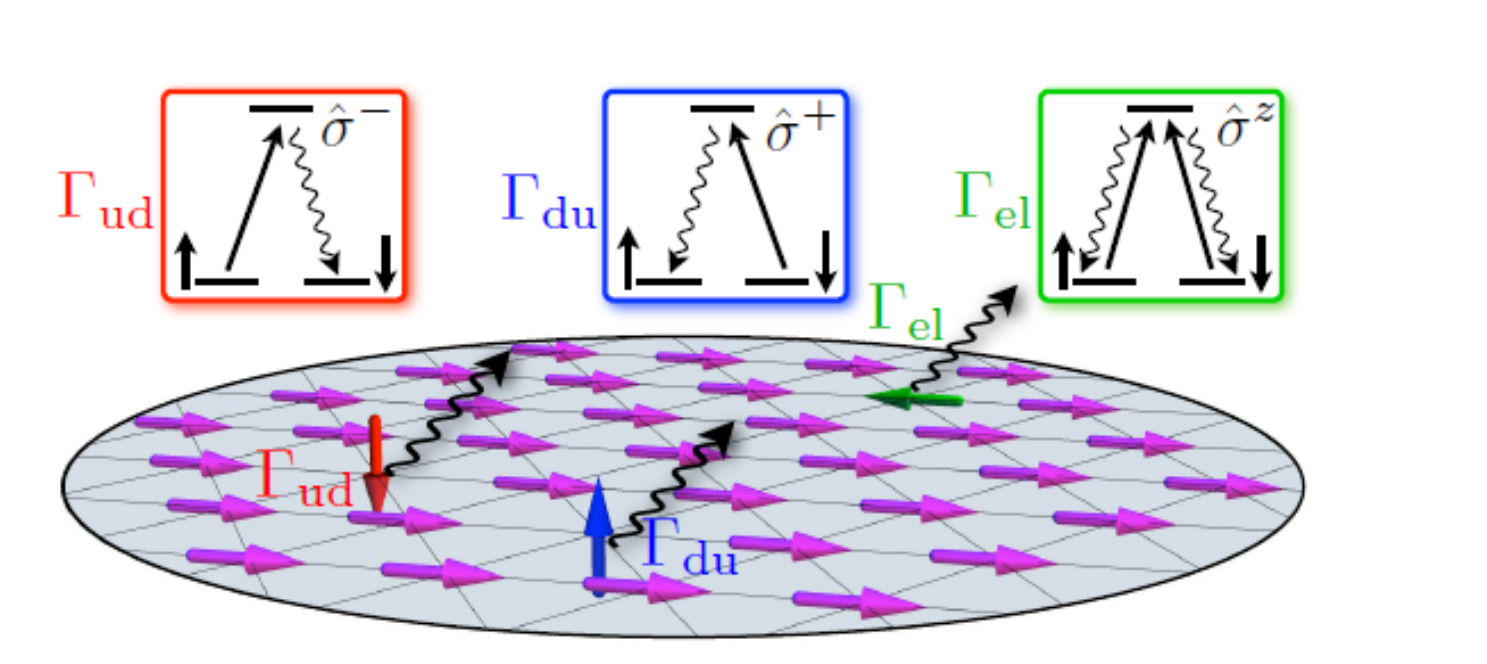}
\caption{Spontaneous emission from the Raman beams creating the spin-dependent force causes three types of decoherence: Raman decoherence processes with rates $\Gamma_{\mathrm{ud}}$ and $\Gamma_{\mathrm{du}}$, which project spins to be down or up, respectively, and elastic decoherence processes, which cause dephasing of a spin superposition state.  Figure from Ref.~\cite{Foss-Feig2013}.}
\label{fig:Fig1}
\end{centering}
\end{figure}

\subsection*{Theoretical Modeling of Spin-spin Interactions with Spontaneous Emission}
As discussed in the earlier section on sources of noise and decoherence, spontaneous emission must be accounted for during the interaction time.  Following Ref.~\cite{Uys2010}, there are three different types of spontaneous emission processes, shown in Fig.~\ref{fig:Fig1}.  The processes with rates $\Gamma_{\mathrm{ud}}$ and $\Gamma_{\mathrm{du}}$, which induce spontaneous transitions from $|\uparrow\rangle$ to $|\downarrow\rangle$ and from $|\downarrow\rangle$ to $|\uparrow\rangle$, respectively, arise from Raman scattering.  On the other hand, Rayleigh scattering, with associated decoherence rate $\Gamma_{\mathrm{el}}$, produces dephasing of a spin superposition state.  In typical experimental realizations, $\Gamma_{\mathrm{ud}},\Gamma_{\mathrm{du}}\sim 10\,$s$^{-1}$, and $\Gamma_{\mathrm{el}}\sim 100-160\,$s$^{-1}$ including the excess decoherence discussed in the above section on spontaneous emission.  The spin dynamics in this case can be modeled by a master equation in Lindblad form, and this master equation admits an exact solution, as discussed in Ref.~\cite{Foss-Feig2013}.  Using this exact solution, we compute the contrast and spin variance expectations from the correlation functions
\begin{align}
\label{eq:s1}\langle \hat{\sigma}^{+}_j\rangle &=\frac{e^{-\Gamma t}}{2}\prod_{k\ne j}\Phi\left(J_{jk},t\right)\, ,\\
\langle \hat{\sigma}^{a}_j\hat{\sigma}^{b}_k\rangle&=\frac{e^{-2\Gamma t}}{4}\prod_{l\notin \{j,k\}}\Phi\left(a J_{jl}+b J_{kl},t\right)\, ,\\
\label{eq:s3}\langle \hat{\sigma}^{a}_j\hat{\sigma}^z_k\rangle&=\frac{e^{-\Gamma t}}{2}\Psi\left(a J_{jk},t\right)\prod_{l\notin \{j,k\}}\Phi\left(a J_{jl},t\right)\, ,
\end{align}
where $a,b\in\{+,-\}$ and
\begin{align}
\nonumber &\Phi\left(J,t\right)=e^{-\frac{\left(\Gamma_{\mathrm{ud}}+\Gamma_{\mathrm{du}}\right) t}{2}}\Big[\cos\left(t\sqrt{\left(2i\gamma+2J/N\right)^2-\Gamma_{\mathrm{ud}}\Gamma_{\mathrm{du}}}\right)\\
&+\frac{\Gamma_{\mathrm{ud}}+\Gamma_{\mathrm{du}}}{2} t\,\sinc\left(t\sqrt{\left(2i\gamma+2J/N\right)^2-\Gamma_{\mathrm{ud}}\Gamma_{\mathrm{du}}}\right)\Big]\, ,\\
\nonumber &\Psi\left(J,t\right)=e^{-\frac{\left(\Gamma_{\mathrm{ud}}+\Gamma_{\mathrm{du}}\right) t}{2}}\left[i\left(2i\gamma+2J/N\right)-2\gamma\right]t\\
&\times \sinc\left(t\sqrt{\left(2i\gamma+2J/N\right)^2-\Gamma_{\mathrm{ud}}\Gamma_{\mathrm{du}}}\right)\, .
\end{align}
In these expressions, the initial state is the product state of all spins pointing along the $x$ direction, $\gamma=\left(\Gamma_{\mathrm{ud}}-\Gamma_{\mathrm{du}}\right)/4$, and $\Gamma=\left(\Gamma_{\mathrm{ud}}+\Gamma_{\mathrm{du}}+\Gamma_{\mathrm{el}}\right)/2$.  In the case that the couplings between spins are uniform, $J_{j,k}=\bar{J}$ for all $j$ and $k$, these results simplify to become
\begin{align}
\label{eq:s1ATA} &\langle \hat{\sigma}^{+}\rangle =\frac{e^{-\Gamma t}}{2}\Phi^{N-1}\left(\bar{J},t\right)\, ,\\
&\langle \hat{\sigma}^{a}\hat{\sigma}^{b}\rangle=\frac{e^{-2\Gamma t}}{4}\Phi^{N-2}\left(\left(a+b\right)\bar{J},t\right)\, ,\\
&\label{eq:s3ATA} \langle \hat{\sigma}^{a}\hat{\sigma}^z\rangle=\frac{e^{-\Gamma t}}{2}\Psi\left(a \bar{J},t\right)\Phi^{N-2}\left(a \bar{J},t\right)\, .
\end{align}

\subsubsection*{Computation of the spin-spin coupling constants}

\begin{figure}
\begin{centering}
\includegraphics[width=0.65\columnwidth]{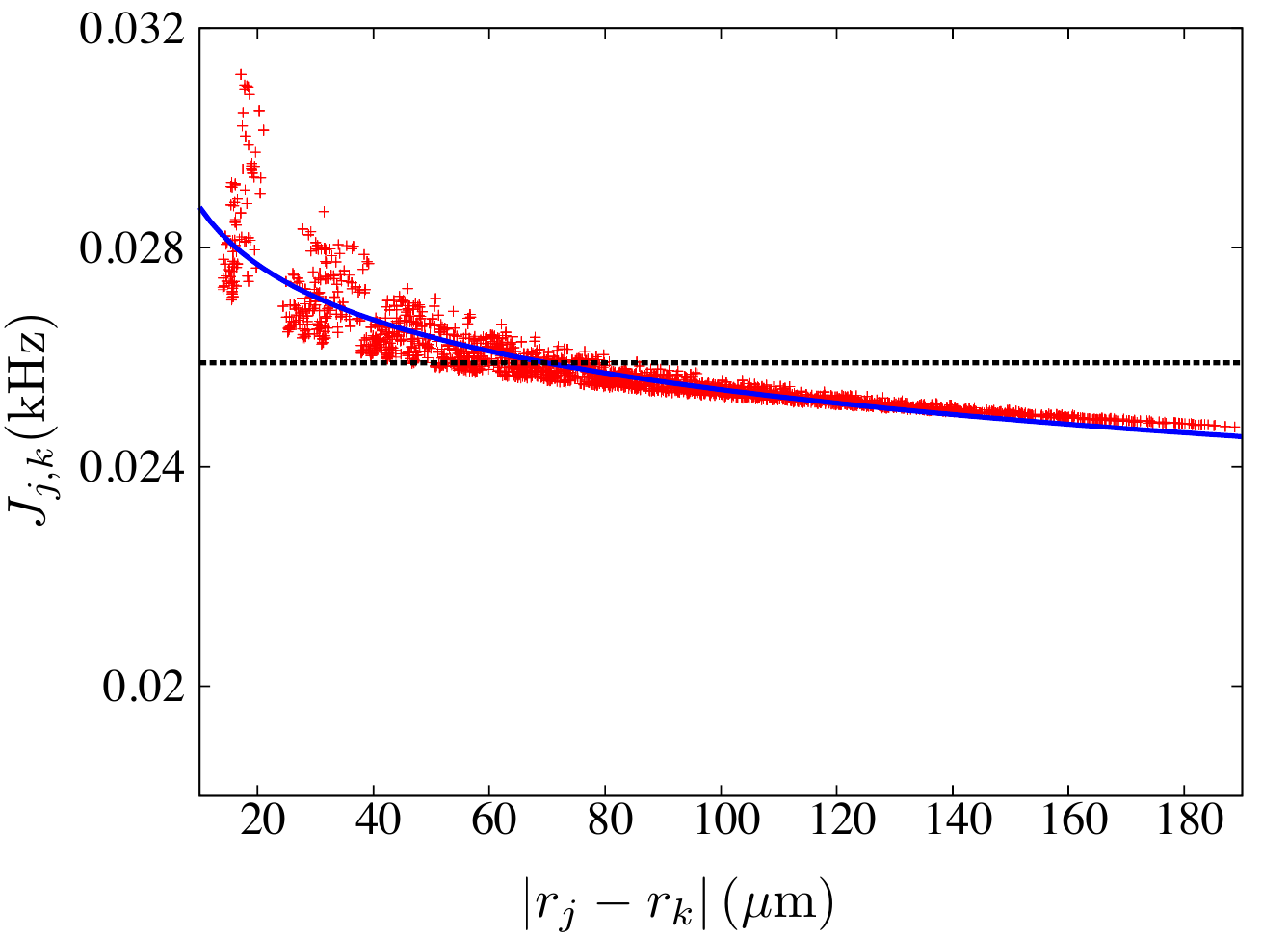}
\caption{Spin-spin couplings for the experimental parameters given in the main text (red points), together with their best power-law fit $\alpha\sim 0.05$  (blue solid line) and the uniform coupling approximation (black dashed line). }
\label{fig:Jijs}
\end{centering}
\end{figure}

We compute the ion crystal equilibrium structure and normal modes numerically following Ref.~\cite{Wang2013}, where it is shown that the spin-spin coupling constants are given by
\begin{align}
\label{eq:spinspsincoup}J_{ij}&=\frac{F_0^2 N}{2\hbar M}\sum_{m=1}^{N}\frac{b_{i,m}b_{j,m}}{\mu^2-\omega_m^2}\, .
\end{align}
Here, $\omega_m$ and $\mathbf{b}_m$ are the frequency and amplitude of normal mode $m$, respectively.  We find that for the experimental parameters used in this work the coupling constants are well-represented by the form $J_{j,k}\propto 1/\left|r_j-r_k\right|^{\alpha}$ with $r_j$ the position of ion $j$ and $\alpha\sim 0.02-0.18$.  As a particular example, the spin-spin coupling constants for $N=127$ ions, $\omega_z=2\pi\times 1.580\,$MHz, a rotating wall rotation frequency of $2\pi\times 180\,$kHz and potential chosen to match experimental mode spectra, and a force and detuning chosen such that $\bar{J}/h=3.30\,$kHz are shown in Fig.~\ref{fig:Jijs}.  A best fit to the computed spin-spin couplings gives a power law $\alpha\sim 0.05$.

\subsubsection*{Validation of the uniform coupling approximation}

\begin{figure}
\begin{centering}
\includegraphics[width=0.65\columnwidth]{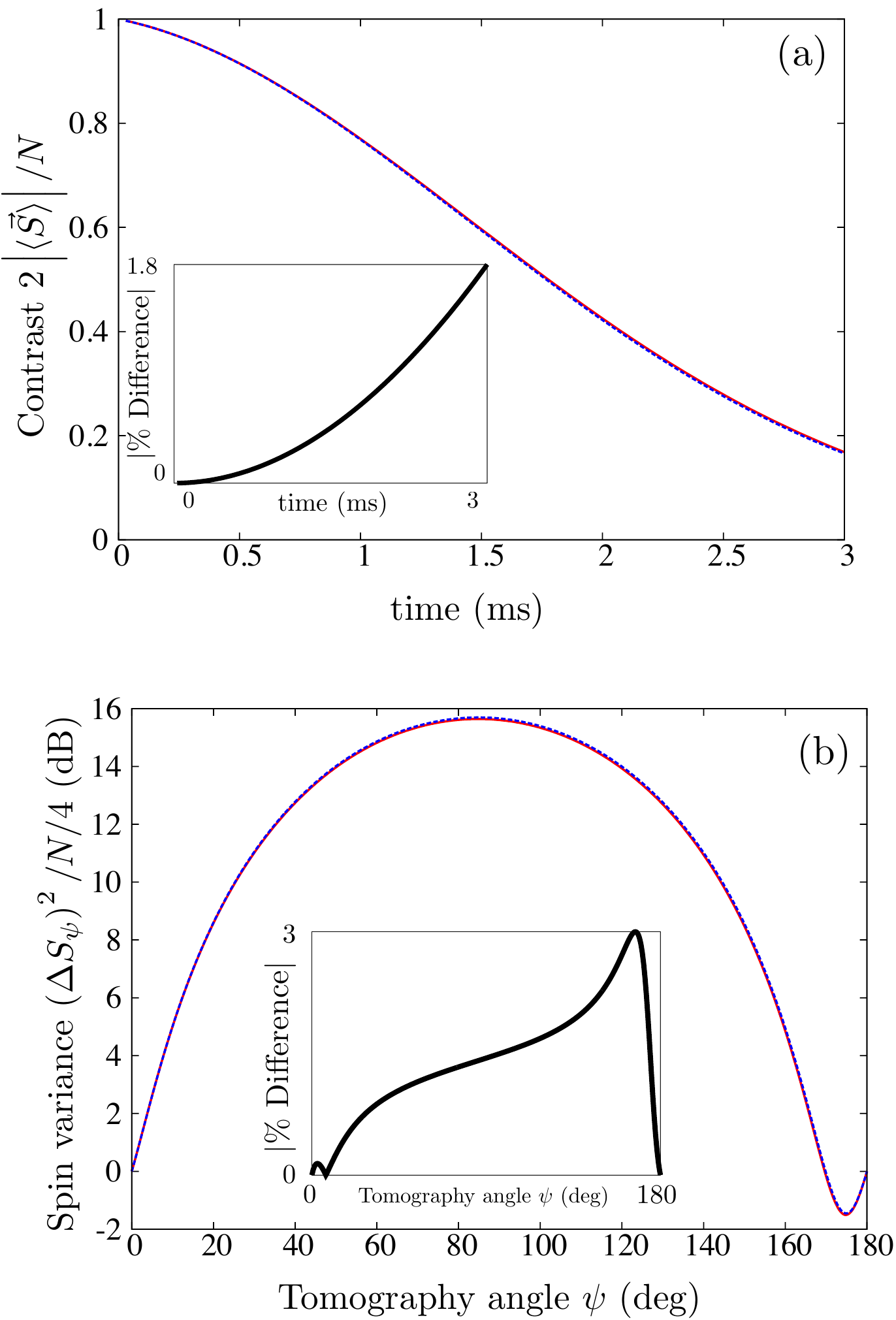}
\caption{Normalized contrast (panel (a)) and spin noise variance (panel (b)) computed using Eqs.~\eqref{eq:s1}-\eqref{eq:s3} (solid red lines) and the uniform coupling approximation Eqs.~\eqref{eq:s1ATA}-\eqref{eq:s3ATA} (dashed blue lines) for the spin-spin couplings of Fig.~\ref{fig:Jijs}.  The insets show the percent difference between the results.  The parameters used are $N=127$, $\bar{J}/h=3.30\,$kHz, $\Gamma_{\mathrm{el}}=171.6$s$^{-1}$, $\Gamma_{\mathrm{ud}}=9.2$s$^{-1}$, $\Gamma_{\mathrm{du}}=6.5$s$^{-1}$, and an interaction time of $3$ms in panel (b). }
\label{fig:ATAComp}
\end{centering}
\end{figure}

Here, we show that the uniform coupling approximation Eqs.~\eqref{eq:s1ATA}-\eqref{eq:s3ATA} used in the main text is a good approximation to the solutions computed using Eqs.~\eqref{eq:s1}-\eqref{eq:s3}, with the spin-spin couplings directly determined  from the phonon modes Eqn.~\eqref{eq:spinspsincoup}.  We compare the two for the parameters of Fig.~\ref{fig:Jijs}.  The decoherence rates are taken to match measured rates of $\Gamma_{\mathrm{el}}=171.6$s$^{-1}$, $\Gamma_{\mathrm{ud}}=9.2$s$^{-1}$, $\Gamma_{\mathrm{du}}=6.5$s$^{-1}$.  The normalized contrast and spin noise variance are compared in Fig.~\ref{fig:ATAComp}(a)-(b) and very little difference is observed. These comparisons  validate the use of the uniform coupling approximation.

\begin{figure}
\begin{centering}
\includegraphics[width=0.65\columnwidth]{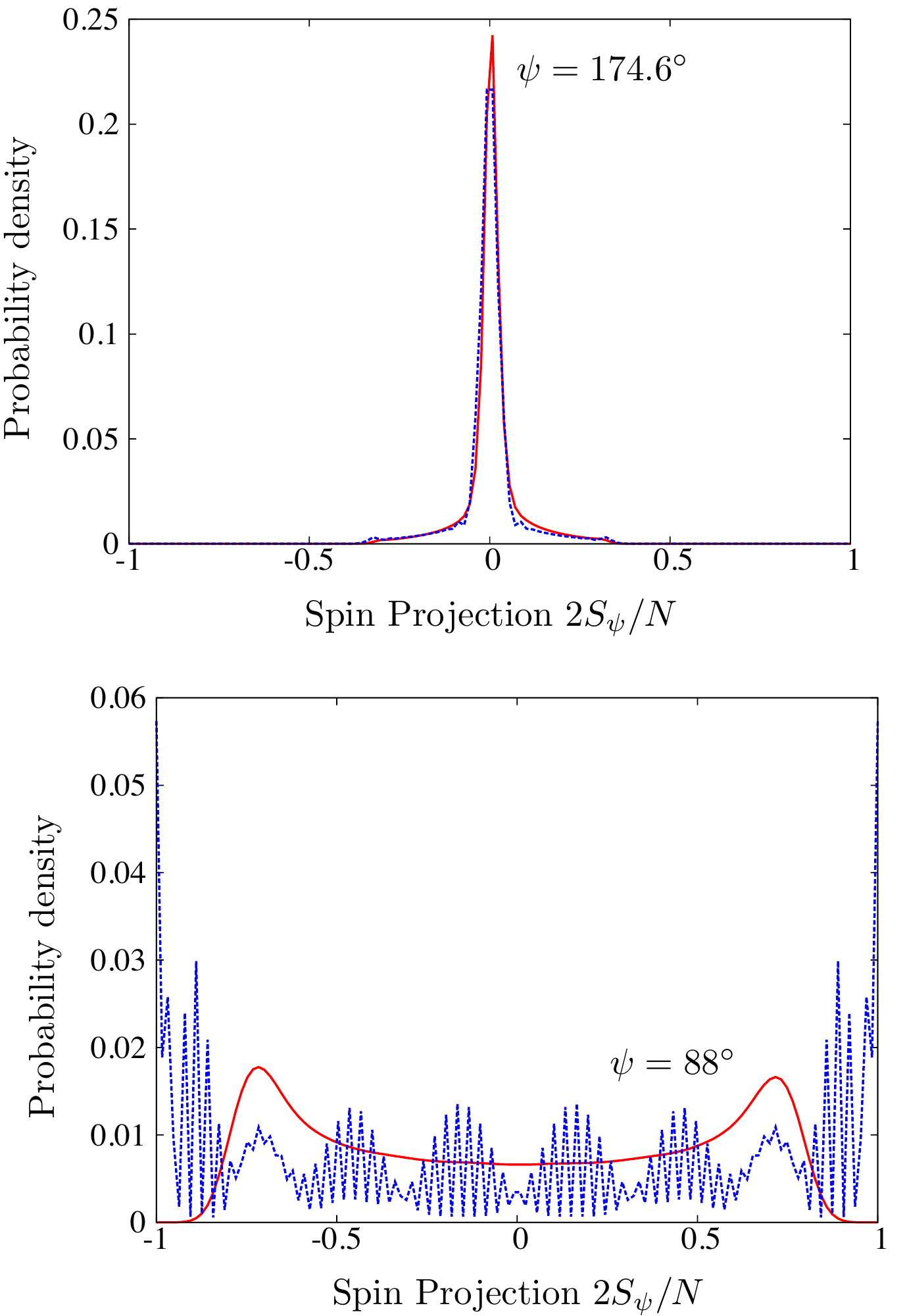}
\caption{Full counting statistics with (red solid) and without (blue dashed) decoherence from spontaneous emission for the squeezed (upper panel, $\psi=174.6^{\circ}$) and anti-squeezed (lower panel, $\psi=88^{\circ}$) quadratures.  Decoherence more significantly affects the anti-squeezed quadrature.}
\label{fig:DCComp}
\end{centering}
\end{figure}

\subsection*{Computation of the Full Counting Statistics}
While a computation of the full counting statistics for general spin-spin couplings $J_{jk}$ is exponentially difficult in the number of ions, the computation can be performed efficiently with the uniform coupling approximation, $J_{jk}=\bar{J}$.  Rather than directly computing the probability distribution to measure $n$ spins along $\hat{S}^{\psi}=\cos\psi \hat{S}^z+\sin\psi \hat{S}^y$, $P_{\psi}\left(n\right)$, it is advantageous to compute the characteristic function
\begin{align}
\label{eq:CharF}{C}_{\psi}\left(q\right)&=\langle e^{i q \sum_{j=1}^{N} \hat{\sigma}^{\psi}_j}\rangle\, .
\end{align}
From the characteristic function,  the probability distribution is obtained by Fourier transformation as
\begin{align}
P_{\psi}\left(n\right)&=\frac{1}{N+1}\sum_{k=0}^{N} e^{-i \frac{2\pi k}{N+1}  n} C_{\psi}\left(\frac{2\pi k}{N+1}\right) \, .
\end{align}
Because the spin operators $\hat{\sigma}^{\psi}_j$ in Eqn.~\eqref{eq:CharF} mutually commute, we can write the characteristic function as
\begin{align}
{C}_{\psi}\left(q\right)&=\langle\prod_{j=1}^{N}  \left[\cos\left(q\right) \hat{I}_j+i\sin\left(q\right) \hat{\sigma}^{\psi}_j\right]\rangle\, ,
\end{align}
where $\hat{I}_j$ is the identity operator for spin $j$.  In the uniform coupling approximation, this becomes
\begin{align}
{C}_{\psi}\left(q\right)&=\langle \left(\cos\left(q\right) \hat{I}+i\sin\left(q\right) \hat{\sigma}^{\psi}\right)^N\rangle \, ,
\end{align}
where it is understood that multiple instances of $\hat{\sigma}^{\psi}$ are interpreted to correspond to operators on different spins.  This result can be expanded using the binomial theorem as
\begin{align}
{C}_{\psi}\left(q\right)&=\sum_{n=0}^{N}\binom{N}{n} \cos^{N-n}\left(q\right)\left(i\sin\left(q\right)\right)^{n} \langle \hat{\sigma}^{\psi}_{\left(n\right)}\rangle \, ,
\end{align}
where $\langle \hat{\sigma}^{\psi}_{\left(n\right)}\rangle$ denotes the expectation with $n$ $\hat{\sigma}^{\psi}$ operators on different spins, e.g., $\langle \hat{\sigma}^{\psi}_{\left(2\right)}\rangle=\langle \hat{\sigma}^{\psi}_1 \hat{\sigma}^{\psi}_2\rangle$.  In the uniform coupling approximation, validated in Fig.~\ref{fig:ATAComp}, the system is permutationally symmetric and so $\langle \hat{\sigma}^{\psi}_i \hat{\sigma}^{\psi}_j\rangle=\langle \hat{\sigma}^{\psi}_{\left(2\right)}\rangle $ for all $i\ne j$.
Expanding a product of $n$ $\hat{\sigma}^{\psi}$ in terms of $n_+$ $\hat{\sigma}^+$s, $n_-$ $\hat{\sigma}^-$s, and $(n-n_+-n_-)$ $\hat{\sigma}^z$s, the final result for the characteristic function in the uniform coupling approximation is
 \begin{align}
\nonumber &\langle C_{\psi}\left(q\right)\rangle=\sum_{n=0}^{N}\binom{N}{n}  \cos^{N-n}\left(q\right)\left(i\sin\left(q\right)\right)^{n} \\
\nonumber &\times \sum_{n_+=0}^{n}\sum_{n_-=0}^{n-n_+}\frac{n!}{n_+!n_-!\left(n-n_+-n_-\right)!}\left(-i\sin\psi\right)^{n_+}\left(i\sin\psi\right)^{n_-}\\
&\times \left(\cos\psi\right)^{n-n_+-n_-}\langle \hat{\sigma}^+_{\left(n_+\right)}\hat{\sigma}^-_{\left(n_-\right)}\hat{\sigma}^z_{\left(n-n_+-n_-\right)}\rangle \, .
\end{align}
Using the methods of Ref.~\cite{Foss-Feig2013}, we can write the correlation function as
\begin{align}
\nonumber& \langle \hat{\sigma}^+_{\left(n_+\right)}\hat{\sigma}^-_{\left(n_-\right)}\hat{\sigma}^z_{\left(n_z\right)}\rangle=\frac{e^{-\left(n_++n_-\right)\Gamma t}}{2^{n_++n_-}} \Psi^{n_z}\left(\left(n_+-n_-\right)J,t\right) \\
&\times \Phi^{N-\left(n_++n_-+n_z\right)}\left(\left(n_+-n_-\right) J,t\right)\, ,
\end{align}
which leads to a fully analytic representation of the counting statistics.  A comparison of the counting statistics with and without the effects of decoherence from spontaneous emission for the parameters of Fig.~\ref{fig:ATAComp} is given in Fig.~\ref{fig:DCComp}.  The quadrature for $\hat{S}^{\psi}$ at $\psi=88^{\circ}$, which is characteristic of antisqueezing, is more strongly affected by decoherence than the quadrature along the squeezed direction, $\psi=174.6^{\circ}$.  The fast oscillations exhibited by the Hamiltonian evolution are washed out by decoherence.

\begin{figure}
\begin{centering}
\includegraphics[width=0.65\columnwidth]{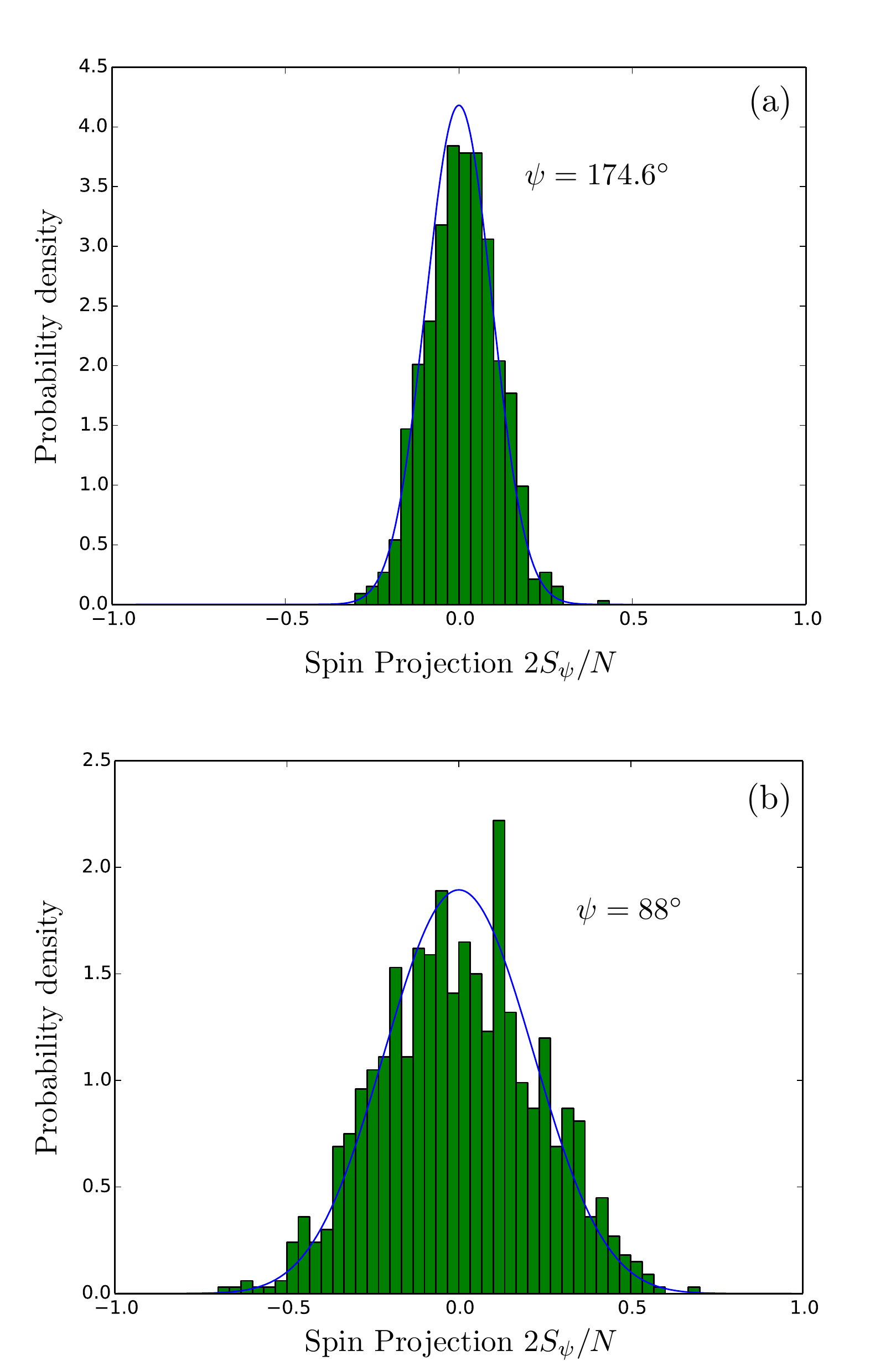}
\caption{Experimentally measured histogram of counting statistics in the absence of ODF beams (bars) compared with the theoretical prediction (solid line) including magnetic field and photon shot noise.  The upper panel is for the squeezed quadrature $\psi=174.6^{\circ}$, while the lower panel is for the anti-squeezed quadrature $\psi=88^{\circ}$.}
\label{fig:NOODF}
\end{centering}
\end{figure}

\subsubsection*{Theoretical modeling of magnetic field fluctuations and photon shot noise}

\begin{figure}
\begin{centering}
\includegraphics[width=0.65\columnwidth]{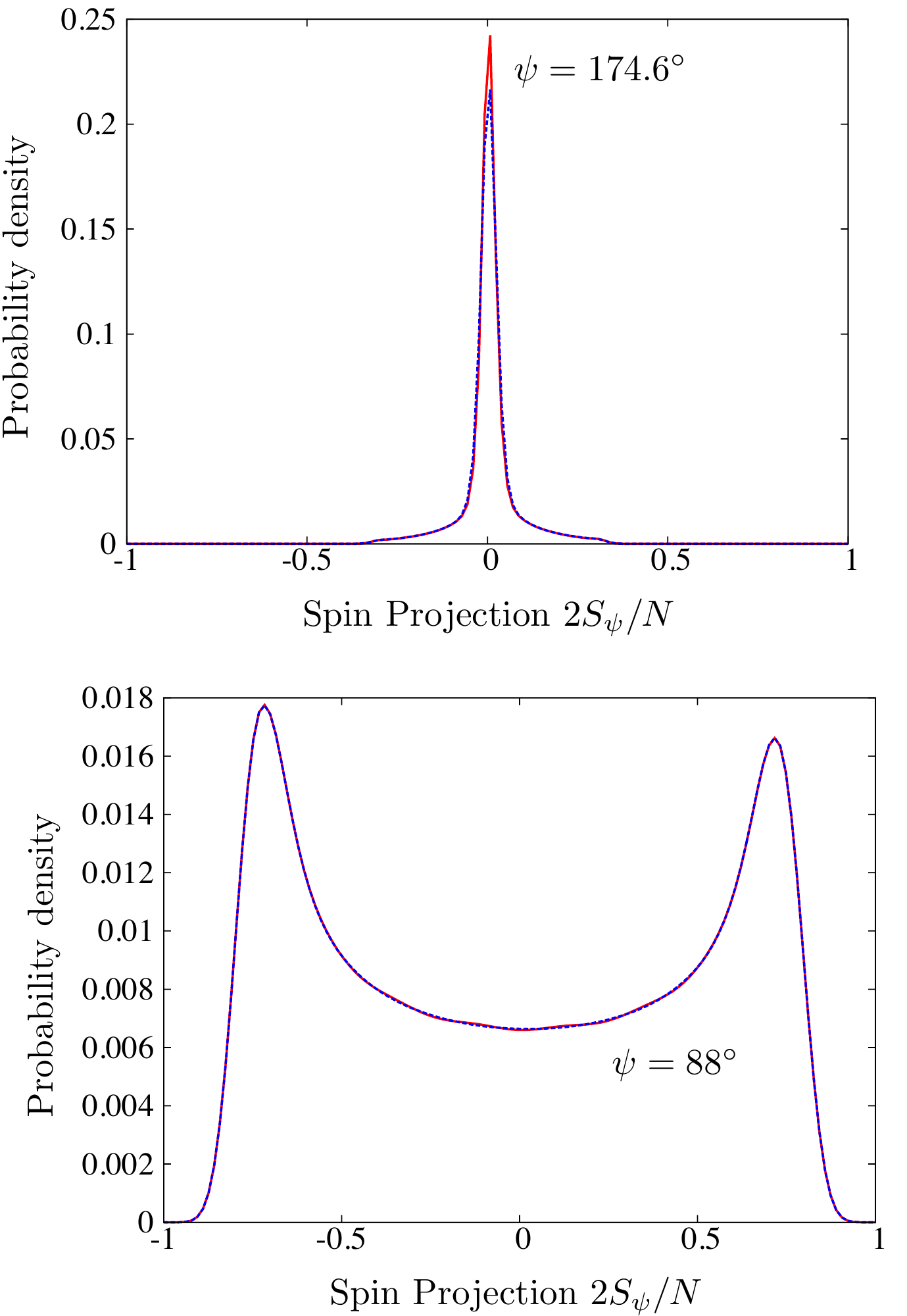}
\caption{Full counting statistics without (red solid) and with (blue dashed) homogeneous magnetic field noise for the squeezed (upper panel, $\psi=174.6^{\circ}$) and anti-squeezed (lower panel, $\psi=88^{\circ}$) quadratures.  The effects of magnetic field noise are strongly suppressed by decoherence. The asymmetry in the peaks at positive and negative $S_{\psi}$ are due to unequal Raman decoherence rates $\Gamma_{\mathrm{du}}\ne \Gamma_{\mathrm{ud}}$.}
\label{fig:BNComp}
\end{centering}
\end{figure}

\begin{figure}
\begin{centering}
\includegraphics[width=0.65\columnwidth]{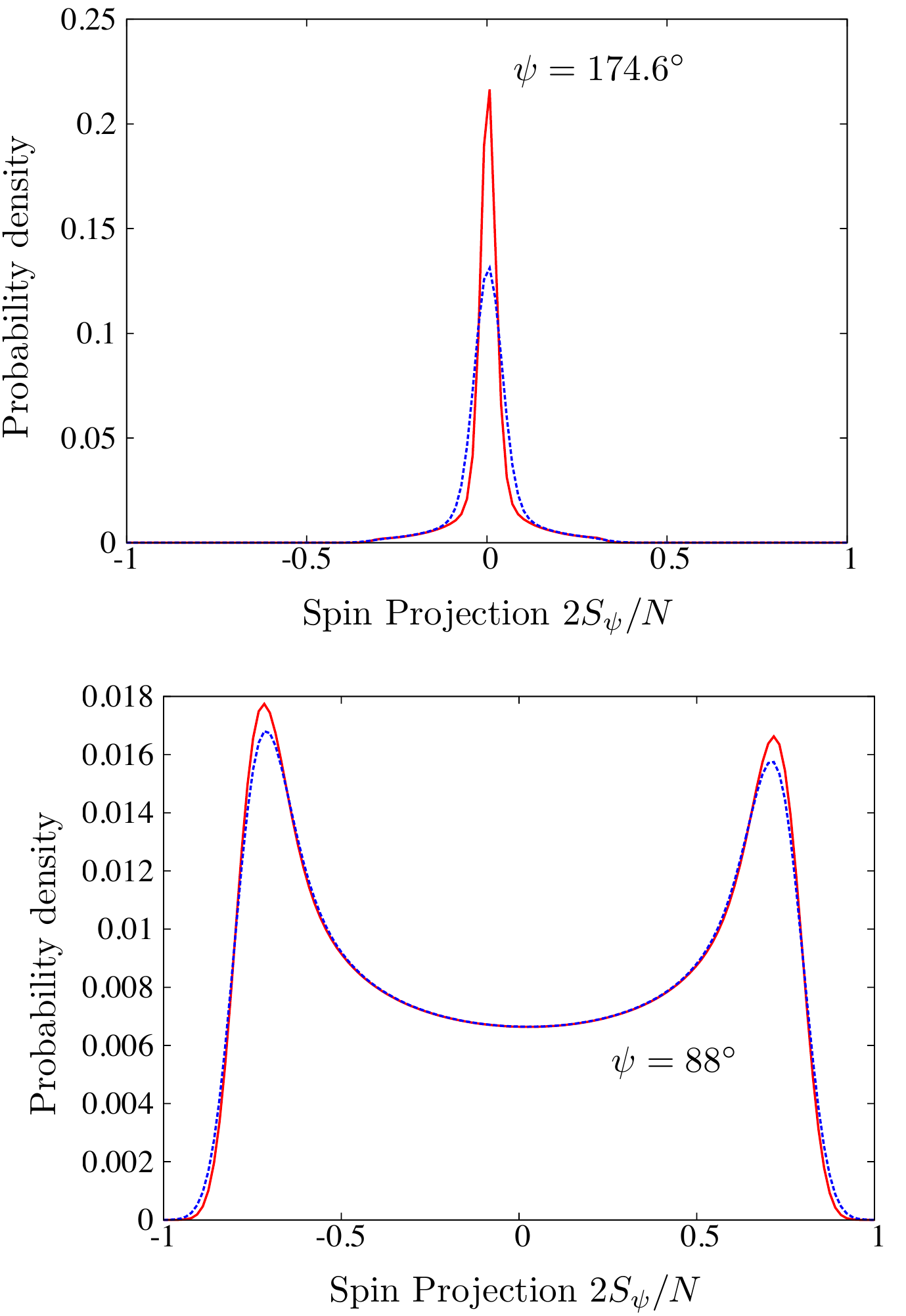}
\caption{Full counting statistics without (red solid) and with (blue dashed) convolution with photon shot noise.  Both curves also include homogeneous magnetic field noise.  The upper panel is for the squeezed quadrature $\psi=174.6^{\circ}$, while the lower panel is for the anti-squeezed quadrature $\psi=88^{\circ}$.  In order to compare the discrete distribution without shot noise to the continuous distribution with shot noise, the latter was evaluated on the support of the former and normalized to sum to 1. Note that the normalization of the data is different than in Fig. 4 of the main text, but the blue lines are otherwise identical.}
\label{fig:ShotNoise}
\end{centering}
\end{figure}

As mentioned in an earlier section, homogenous fluctuations in the magnetic field caused by vibrations of the magnet contribute to dephasing.  In the absence of decoherence, the effect of a time-fluctuating, homogeneous magnetic field with Hamiltonian $\hat{H}_{B}=B\left(t\right)\sum_{i}\hat{S}^z_i$ on an arbitrary permutation-symmetric correlation function is
\begin{align}
\langle \hat{\sigma}^+_{\left(n_+\right)}\hat{\sigma}^-_{\left(n_-\right)}\hat{\sigma}^z_{\left(n_z\right)}\rangle_B&=e^{i\varphi\left(\tau\right)\left(n_+-n_-\right)}\langle \hat{\sigma}^+_{\left(n_+\right)}\hat{\sigma}^-_{\left(n_-\right)}\hat{\sigma}^z_{\left(n_z\right)}\rangle_{B\to 0}\, ,
\end{align}
where $\varphi\left(\tau\right)=\int_0^{\tau/2} dt B\left(t\right)-\int_{\tau/2}^{\tau} dt B\left(t\right)$ for the spin-echo sequence used in the experiment.  While this expression is no longer exact in the presence of Raman decoherence, it is an excellent approximation in the experimentally relevant case that $B\left(t\right)$ is small and elastic decoherence dominates over Raman decoherence.  Averaging over realizations of the fluctuating field $\varphi\left(\tau\right)$, we find
\begin{align}
\nonumber \overline{e^{i\varphi\left(\tau\right)\left(n_+-n_-\right)}}&= 1-\frac{\left(n_+-n_-\right)^2}{2}\Delta \phi_{rms}^2\left(\tau\right)+\dots\\
&= e^{-\frac{\left(n_+-n_-\right)^2}{2}\Delta \phi_{rms}^2\left(\tau\right)}\, ,
\end{align}
where the overbar denotes averaging of the stochastic variable, $\Delta \phi_{rms}^2\left(\tau\right)=\overline{\varphi^2\left(\tau\right)}$, and we have used the fact that $\overline{\varphi\left(\tau\right)}=0$. The variance $\Delta \phi_{rms}^2\left(\tau\right)$ is determined experimentally, as shown in Fig.~\ref{Bfield fluctuations}.  With this, we find that the correlation functions in the presence of a fluctuating magnetic field are obtained as
\begin{align}
&\label{eq:TheoryBfluc}\overline{\langle \hat{\sigma}^+_{\left(n_+\right)}\hat{\sigma}^-_{\left(n_-\right)}\hat{\sigma}^z_{\left(n_z\right)}\rangle_B}\\
\nonumber &\approx e^{-\frac{\left(n_+-n_-\right)^2\Delta \phi_{rms}^2\left(\tau\right)}{2} }\langle \hat{\sigma}^+_{\left(n_+\right)}\hat{\sigma}^-_{\left(n_-\right)}\hat{\sigma}^z_{\left(n_z\right)}\rangle_{B\to 0}\, .
\end{align}
A comparison of the experimentally measured counting statistics with no ODF beams ($\bar{J}$ and all decoherence rates set to zero) with the theory that models magnetic field noise with Eqn.~\eqref{eq:TheoryBfluc} is shown in Fig.~\ref{fig:NOODF}, demonstrating that the noise is accounted for consistently.  In particular, we see that the dependence on tomography angle $\psi$ of the variance due to magnetic field noise is accurately captured.    In Fig.~\ref{fig:BNComp}, we show the difference between the theoretical predictions for the counting statistics with and without homogeneous magnetic field noise, using the parameters of Fig.~\ref{fig:ATAComp}, a total interaction time of $\tau=3$ms, and $\Delta \phi_{rms}^2\left(\tau\right)=0.035$.  We see that the magnetic field noise has a relatively slight effect on the full counting statistics.  This can be understood by noting that the correlations which are significantly affected by the magnetic field noise (those with $(n_+-n_-)\ne 0$) are already suppressed by the factor $e^{- \left(n_++n_-\right) \Gamma t}$ due to decoherence from spontaneous emission.  The final source of noise we include in our theoretical predictions is photon shot noise, which is accounted for by convolving the theoretical counting statistics ${P}_{\psi}$ with the distribution of photon shot noise $P_{\mathrm{psn}}$.  For the present case, the distribution of shot noise is taken to be Gaussian with standard deviation $0.03$ in units of $S_{\psi}/(N/2)$ (11\% of spin projection noise).  A comparison of the results with and without this convolution is given in Fig.~\ref{fig:ShotNoise}. Also, we note that the dashed lines in Fig.~\ref{fig:ShotNoise}, which include all the sources of noise described above, are plotted in Fig.~4 of the main text with a different normalization. In order to compare the discrete probability distribution which does not include shot noise with the continuous distribution resulting from the convolution, the latter is evaluated on the set of points where the former has support and normalized so that its sum is unity.

\subsubsection*{Extraction of the Fisher information from the Hellinger distance}

The Fisher information $F$, which measures the distinguishability of quantum states with respect to small phase rotations, is a many-particle entanglement witness, with a measurement of $F/N>n$ implying that the state is $n$-particle entangled.  Importantly, this characterization of entanglement holds even for non-Gaussian states, where spin squeezing is no longer an effective witness.  Further, the bound $F/N>1$ is both a necessary and sufficient condition for the ability to perform sub-shot-noise phase estimation with a quantum state.

\begin{figure}
\begin{centering}
\includegraphics[width=0.65\columnwidth]{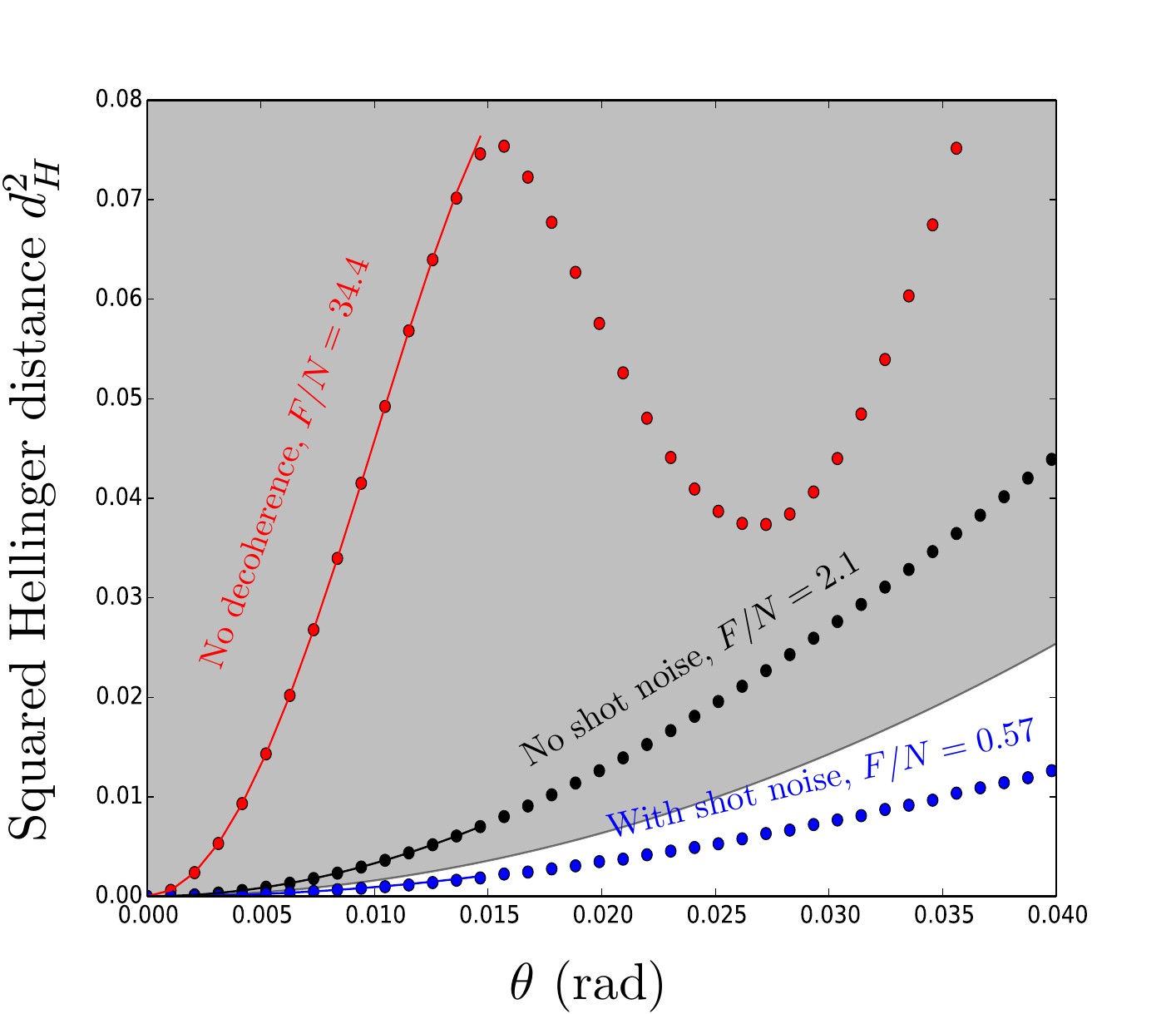}
\caption{Extraction of the Fisher information from the theoretically computed Hellinger distance without (black) and with (blue) photon shot noise, including the effects of decoherence from spontaneous emission and magnetic field noise.  Also shown in red is the Hellinger distance in the absence of decoherence or photon shot noise, for comparison.  The points denote computed values of the Hellinger distance and the lines are small-angle quartic fits used to extract the Fisher information.  The gray swath denotes the region of entangled states.}
\label{fig:FisherTheory}
\end{centering}
\end{figure}

We characterize the Fisher information following the method of Ref.~\cite{Strobel2014}, which utilizes the Euclidean distance in the space of probability amplitudes known as the (squared) Hellinger distance,
\begin{align}
d_H^2\left(\theta\right)&=\frac{1}{2}\sum_n\left(\sqrt{P_{\theta}\left(n\right)}-\sqrt{P_{0}\left(n\right)}\right)^2\, .
\end{align}
Here, $P_{0}\left(n\right)$ denotes the counting statistics along the optimal tomography angle $\psi$, $P_{\theta}\left(n\right)$ denotes the counting statistics after rotation by $\theta$ around $y$, and $\sum_n$ denotes the metric such that $\sum_nP_{\theta}\left(n\right)=1$.  As shown in Ref.~\cite{Strobel2014}, for small angles $\theta$, the squared Hellinger distance satisfies
\begin{align}
d_H^2\left(\theta\right)&=\frac{F}{8}\theta^2+\mathcal{O}\left(\theta^3\right)\, .
\end{align}
We use a quartic fit to the squared Hellinger distance for small rotation angles to extract the quadratic coefficient, and from this the Fisher information per particle.  In the absence of photon shot noise, but including the decoherence from spontaneous emission and magnetic field noise, we find $F/N=2.1$, while including photon shot noise drops this value to $F/N=0.57$, as shown in Fig.~\ref{fig:FisherTheory}.  This comparison demonstrates that the experiment measures an over-squeezed state consistent with entanglement, but that this entanglement cannot be verified with the present magnitude of photon shot noise.  For comparison, we also show the result in the absence of noise or decoherence of any kind, in which case the Fisher information per particle is $F/N=34.4$.

